\documentclass[12pt]{article}
\usepackage{endnotes}
  
\usepackage{graphicx,wrapfig,lipsum}
\usepackage{epstopdf}
\usepackage{amsthm}
\usepackage{amsmath}
\usepackage{amssymb}
\usepackage{amstext}
\usepackage{amsfonts}
\usepackage{enumerate}
\usepackage[authoryear,nonamebreak,bibstyle]{natbib}
\usepackage{footnote}
\makesavenoteenv{tabular}
\makesavenoteenv{table}
\usepackage[onehalfspacing]{setspace}
\usepackage[shortlabels]{enumitem}
\usepackage{multirow}
\usepackage[title]{appendix}
\usepackage{xcolor}
\usepackage{cancel}
\usepackage{hyperref}
\usepackage[normalem]{ulem}
\usepackage{tikz}
\usetikzlibrary{shapes,arrows}
\usepackage{tikz-cd}

\usepackage{comment}

\newif\ifendnote

\begin{document}
\baselineskip=.22in\parindent=30pt

\newtheorem{tm}{Theorem}
\newtheorem{dfn}{Definition}
\newtheorem{lma}{Lemma}
\newtheorem{assu}{Assumption}
\newtheorem{prop}{Proposition}
\newtheorem{cro}{Corollary}
\newtheorem*{theorem*}{Theorem}
\newtheorem{example}{Example}
\newtheorem{observation}{Observation}
\newcommand{\exm}{\begin{example}}
\newcommand{\exmm}{\end{example}}
\newcommand{\obs}{\begin{observation}}
\newcommand{\obss}{\end{observation}}
\newcommand{\cor}{\begin{cro}}
\newcommand{\corr}{\end{cro}}
\newtheorem{exa}{Example}
\newcommand{\ex}{\begin{exa}}
\newcommand{\exx}{\end{exa}}
\newtheorem{remak}{Remark}
\newcommand{\rmk}{\begin{remak}}
\newcommand{\rmkk}{\end{remak}}
\newcommand{\thm}{\begin{tm}}
\newcommand{\nt}{\noindent}
\newcommand{\thmm}{\end{tm}}
\newcommand{\lm}{\begin{lma}}
\newcommand{\lmm}{\end{lma}}
\newcommand{\ass}{\begin{assu}}
\newcommand{\asss}{\end{assu}}
\newcommand{\df}{\begin{dfn}  }
\newcommand{\dff}{\end{dfn}}
\newcommand{\prp}{\begin{prop}}
\newcommand{\prpp}{\end{prop}}
\newcommand{\bqu}{\sloppy \small \begin{quote}}
\newcommand{\equ}{\end{quote} \sloppy \large}
\newcommand\cites[1]{\citeauthor{#1}'s\ (\citeyear{#1})}

\newcommand{\eq}{\begin{equation}}
\newcommand{\eqq}{\end{equation}}
\newtheorem{claim}{\it Claim}
\newcommand{\cl}{\begin{claim}}
\newcommand{\cll}{\end{claim}}
\newcommand{\bit}{\begin{itemize}}
\newcommand{\eit}{\end{itemize}}
\newcommand{\ben}{\begin{enumerate}}
\newcommand{\een}{\end{enumerate}}
\newcommand{\bcen}{\begin{center}}
\newcommand{\ecen}{\end{center}}
\newcommand{\fn}{\footnote}
\newcommand{\ds}{\begin{description}}
\newcommand{\dss}{\end{description}}
\newcommand{\prf}{\begin{proof}}
\newcommand{\prff}{\end{proof}}
\newcommand{\cs}{\begin{cases}}
\newcommand{\css}{\end{cases}}
\newcommand{\ml}{\item}
\newcommand{\lb}{\label}
\newcommand{\ra}{\rightarrow}
\newcommand{\tra}{\twoheadrightarrow}
\newcommand*{\supp}{\operatornamewithlimits{sup}\limits}
\newcommand*{\inff}{\operatornamewithlimits{inf}\limits}
\newcommand{\nf}{\normalfont}
\renewcommand{\Re}{\mathbb{R}}
\newcommand*{\mmax}{\operatornamewithlimits{max}\limits}
\newcommand*{\mmin}{\operatornamewithlimits{min}\limits}
\newcommand*{\argmax}{\operatornamewithlimits{arg max}\limits}
\newcommand*{\argmin}{\operatornamewithlimits{arg min}\limits}
\newcommand{\uhr}{\!\! \upharpoonright  \!\! }

\newcommand{\CR}{\mathcal R}
\newcommand{\CC}{\mathcal C}
\newcommand{\CT}{\mathcal T}
\newcommand{\CS}{\mathcal S}
\newcommand{\CM}{\mathcal M}
\newcommand{\CL}{\mathcal L}
\newcommand{\CP}{\mathcal P}
\newcommand{\CN}{\mathcal N}

\newtheorem{innercustomthm}{Theorem}
\newenvironment{customthm}[1]
  {\renewcommand\theinnercustomthm{#1}\innercustomthm}
  {\endinnercustomthm}
\newtheorem{einnercustomthm}{Extended Theorem}
\newenvironment{ecustomthm}[1]
  {\renewcommand\theeinnercustomthm{#1}\einnercustomthm}
  {\endeinnercustomthm}
  
  \newtheorem{innercustomcor}{Corollary}
\newenvironment{customcor}[1]
  {\renewcommand\theinnercustomcor{#1}\innercustomcor}
  {\endinnercustomcor}
\newtheorem{einnercustomcor}{Extended Theorem}
\newenvironment{ecustomcor}[1]
  {\renewcommand\theeinnercustomcor{#1}\einnercustomcor}
  {\endeinnercustomcor}
    \newtheorem{innercustomlm}{Lemma}
\newenvironment{customlm}[1]
  {\renewcommand\theinnercustomlm{#1}\innercustomlm}
  {\endinnercustomlm}

\newcommand{\red}{\textcolor{red}}
\newcommand{\blue}{\textcolor{blue}}
\newcommand{\purple}{\textcolor{purple}}
\newcommand{\mred}[1]{\color{red}{#1}\color{black}}
\newcommand{\mblue}[1]{\color{blue}{#1}\color{black}}
\newcommand{\mpurple}[1]{\color{purple}{#1}\color{black}}

\renewcommand{\refname}{\large{B~~~Other References}}

\makeatletter
\newcommand{\customlabel}[2]{%
\protected@write \@auxout {}{\string \newlabel {#1}{{#2}{}}}}
\makeatother


\def\qed{\hfill\vrule height4pt width4pt
depth0pt}
\def\reff #1\par{\noindent\hangindent =\parindent
\hangafter =1 #1\par}
\def\title #1{\begin{center}
{\Large {\bf #1}}
\end{center}}
\def\author #1{\begin{center} {\large #1}
\end{center}}
\def\date #1{\centerline {\large #1}}
\def\place #1{\begin{center}{\large #1}
\end{center}}

\def\date #1{\centerline {\large #1}}
\def\place #1{\begin{center}{\large #1}\end{center}}
\def\intr #1{\stackrel {\circ}{#1}}
\def\R{{\rm I\kern-1.7pt R}}
 \def\N{{\rm I}\hskip-.13em{\rm N}}
 \newcommand{\cprod}{\Pi_{i=1}^\ell}
\let\Large=\large
\let\large=\normalsize


\begin{titlepage}
\def\thefootnote{\fnsymbol{footnote}}
\vspace*{0.05in}

\title{The Yannelis--Prabhakar  Theorem on Upper Semi-Continuous \\ \vskip .7em    Selections in   Paracompact Spaces: Extensions and  Applications\fn{Some of the results reported here were first presented by Khan at the {\it Summer Workshop in Economic Theory (SWET)} on October 25, 2018 under the title \lq\lq Nicholas Yannelis and Equilibrium
Theory: Salient Contributions to Economics and Mathematics." He thanks Bernard Cornet, Ed Prescott and Anne Villamil for discussion and encouragement at his talk; he should also like to acknowledge Greg Duffie and his JHU colleagues for a departmental discussion on proper names in connection with the  renaming of the departmental Ely Lectures. This final submission has benefited substantially 
from stimulating conversation and correspondence  with  Youcef Askoura, Marcus Berliant, Kalyan Chatterjee, Ying Chen, Marcelo Ariel Fernandez, Aniruddha Ghosh, Rich McLean and Nicholas Yannelis: the authors thank them all, and also record their gratitude to  Rich McLean  for  his 2020 Ischia slides for giving the first draft a careful reading.  Final thanks to Nicholas Yannelis for sharing the final version of his 1983 Rochester Ph.D. dissertation with the authors.    } } 


\vspace*{0pt}
 
\author{M. Ali Khan\fn{Department of Economics, Johns Hopkins University, Baltimore, MD 21218 {\bf Email:} akhan@jhu.edu} and  Metin Uyan{\i}k\footnote{{\it Corresponding author:} School of Economics, University of Queensland, Brisbane, QLD 4072 \\ {\bf Email:} m.uyanik@uq.edu.au; Orcid ID: 0000-0003-0224-7851}}

%

\vspace*{8pt}

\date{June 27, 2020}


\vskip 1.00em

\bigskip

\bigskip

\baselineskip=.18in

\noindent{\bf Abstract:}  In a 1983 paper, Yannelis--Prabhakar rely on  Michael's selection theorem to guarantee a   continuous selection in the context of the existence of maximal elements and equilibria in abstract economies.  In this tribute to Nicholas Yannelis, we  root this paper in  Chapter II of Yannelis' 1983 Rochester Ph.D. dissertation, and identify its pioneering application of  the {\it paracompactness} condition to current and ongoing work of Yannelis and his co-authors, and to  mathematical economics more generally. We  move beyond the literature  to provide   a necessary and sufficient condition for upper semi-continuous local and global selections of correspondences, and to provide application to   five domains  of Yannelis' interests:
 Berge's maximum theorem, the Gale-Nikaido-Debreu  lemma, the Gale-McKenzie survival assumption, Shafer's non-transitive setting, and the Anderson-Khan-Rashid  approximate existence  theorem. The last resonates with  Chapter VI of the Yannelis' dissertation.  
\hfill  (136~words)


\vskip 5em

\noindent {\it 2010 Mathematics Subject} Classification Numbers: 91B55, 37E05.

\vskip 0.8em

\noindent {\it Journal of Economic Literature} Classification
Numbers: C00, D00, D01

\vskip 0.8em


\vskip 0.8em

\noindent {\it Running Title:}  YP-Theorem: Extensions and  Applications

\vskip 2em

\end{titlepage}

\bigskip

\bigskip

\tableofcontents

\pagebreak

\setcounter{footnote}{0}



\setlength{\abovedisplayskip}{0.1cm}
\setlength{\belowdisplayskip}{0.1cm}

\ifendnote
\let\footnote=\endnote
\fi

\bqu {\it He was one of the young founders ...  He has always been on the side of angels. He has not sold out to the interests nor bought in to romantic idiocies. He stands up to speak for the public interest as his scholarly findings interpret that interest, letting those who will charm college Deans.\fn{Samuelson is referring to $ReStud,$  adding that  Lerner's  \lq\lq contributions as much as anyone's made its first volumes so exciting."  Without overstating the parallel,  Yannelis was one of the young founders of  {\it Economic Theory} along with Aliprantis, Khan, Majumdar and Cass. 
 In a portrait that relies on, and revolves around, Yannelis' first publication,  Samuelson's continuation is  also relevant:  \lq\lq You can gauge the quality of a scientist by his first paper. When the editors asked me to sing of the man and his work, I was delighted. For like Tom Sawyer, who enjoyed his own funeral, I believe the best wakes to be those in which the guest of honour  is present in the full vigour of his powers." 
See Chapter (3:183) referring to Chapter 183 of Volume 3) in the  {\it  Collected Papers;} \cite{sa-c}.   Throughout this work, we conform to this  convention in quotations, and  shall  be relying on on the individual author's own italicizations.  }}   \hfill{Paul Samuelson (1964) on Abba Lerner}  \equ 


\bqu{\it A fresh survey ...  requires the reverse of a
chronological order.  The 
model provides by itself a prism with which to
diffract the paradigms of  various
brands of neoclassicism; and, self-reflexively,
it can serve to help judge the author;  the    
  unity of whose scientific vision  then becomes
 visible.\fn{Samuelson is referring  Sraffa's  {\it magnum opus} \lq\lq The Production of Commodities by Means of Commodities." He continues, \lq\lq  First comes his 1960 book,
which has spawned an extensive literature but still
needs --  if the technology is to be adequately
handled --  to have Sraffa's special equalities
embedded in the general inequalities--equalities of
the 1937 von Neumann model." The model is the \lq\lq essentially
completed Sraffa--Leontief circulating capital
model;" the others are Marx and Ricardo; and Sraffa himself is sighted, along with Dobb,  as  Ricardo's editor; see (6: 411),  also (6:410 and 416).} } \hfill{Paul Samuelson (1987) on Piero Sraffa}\equ 

\bqu {\it When we turn to take a retrospective view, we are always looking and looking away at the same time.\fn{See page ix of \cites{se03} \lq\lq  Natural History of Destruction."}}   \hfill{W. G. Sebald (2003)}  \equ

\smallskip 

\section{Introduction} \lb{sec: introduction} 


\setlength{\abovedisplayskip}{0.1cm}
\setlength{\belowdisplayskip}{0.1cm}



With the turn of political economy to economic science, portraiture has certainly gone out of fashion -- Paul Samuelson may well have been the last great portraitist of our times.\fn{In addition to his portraits of Lerner, Tsuru (3;297) and Solow (7:  534),   see Part XVI in Volume 2, Part XVI in Volume 3, Part X in Volume 4, Parts II and VII in Volume 5, Part II in Volume 6 and Parts IX and X in Volume 7.  Keynes' {\it Essays in Biography} constitute another worthy exemplar; see \cite{ke33}.}     A portrait enables an overview of  an scholar's {\it oeuvre,}  determines  what is  worth giving salience to,  and what needs to be sidelined in a particular work,  parts that need darker colours and those that  could do with  shading out; but the professional use of the portrait extends beyond the one portrayed, and  the portrayer(s),  not to speak of the art of portraying itself. A portrait sheds important light on the profession and the professional understanding of the subject  at the time that it  is being drawn.  As such, it is useful if well-done and if it does not descend into hagiography.\fn{Samuelson (2007) and \cite{st65}, among others,   also wrote on this issue; for the former, see Chapter (527, Volume 7)  titled \lq\lq Reflections on How Biographies of Individual Scholars can Relate to a Science's Biography." As to the latter, see \cite{ks20} for a view that strongly dissents from that of Stigler.}        Even if the talents and gifts of Paul Samuelson are out of our reach, the authors 
 feel that one ought nevertheless to try.  This is one such attempt: a portrait of the mathematical economist Nicholas C. Yannelis done for his sixty-fifth birthday, surely with affection and admiration, but also with some measured distance. It is in keeping with  an eye to the development of  mathematical economics  since Nicholas Yannelis' wrote his 1983 Ph.D. dissertation, and his influential  1983 paper with N. Prabhakar, (still somewhat neglected), down  to the present as encapsulated in his 2016 paper with Wei He.

With this preamble then out of the way, we can get to work.  In order to attain some sense of perspective,\fn{We use the term in a technical sense, as in \cite{ke78}, and   as we shall have occasion to emphasize in the concluding Section 4, we exclude Yannelis' more recent work on the economics of information. Kemp writes in the context of linear perspective \lq\lq  [O]ne of the primary characteristics of linear perspective is that it describes precisely
the point at which one object occludes part of another more distant form (p.138)." \label{fn:ke}}  the focus of this work  shall be  on the  Yannelis-Prabhakar  selection theorem for lower semi-continuous correspondences  in the setting of  paracompact spaces --  it shall   chart   what the authors see to be progress in the subject of equilibrium existence theory in its Walrasian and Cournotian-Nash modes  since this paper was written roughly four decades ago.  To be sure, the paper already had its echo\fn{The ordering of the authors' names already alludes to this.}    in the second chapter of Yannelis' dissertation, his debut in the public domain, a chapter   written in the light and the shade of Mas-Colell's 1974  dramatic announcement that for  \lq\lq the  general equilibrium Walrasian model to be well-defined and consistent (i.e., for it to have a solution), the hypotheses of completeness and transitivity of consumer's preferences are not needed." His paper, supplemented by the efforts of Gale, Shafer and Sonnenschein\fn{The relevant references are  \cite{ma74}, along with \cite{gm75, gm79} and \cite{sh74, sh76} and \cite{ss75}.} served as the leitmotiv not only for the thesis but also for Yannelis'  work for the major part of the subsequent decade, and still showing traces in his work.  

Since our primary focus is on \cite{yp83} (henceforth YP-Paper), and since we see it to be  rooted  in   Chapter 1 of the dissertation (henceforth Y-Chapter),   we use the remainder of this introduction to outline the chapter, and to frame it in  the dissertation as a whole. Towards this end, we  begin with a summary statement that the  ambitious and synthetic reach of the chapter is already  itself  evident in its single sentence introduction: 

\bqu The purpose of this chapter is to develop mathematical tools which can be applied to prove in a very simple and general way the nonemptiness of demand sets, the existence of competitive equilibrium and the existence of a Nash equilibrium.    \equ

\nt The chapter  relies on Michael's selection theorem invoked by\fn{It is worth noting what \cite{gm79} say in this context: \lq\lq The proof needs no modification because the selection theorem of Michael which is used requires only a lower semi-continuous correspondence;" also see \cite{gm77}.  We shall return to this issue of the proof  in the sequel. \label{fn:gm}}  \cite{gm75}, and on Fan's 1961 generalization of the Knaster-Kuratowski-Mazurkiewicz (KKM) lemma, again invoked in mathematical economics  by\fn{Gale writes with reference to the 1954 paper of Arrow-Debreu, \lq\lq  However, where the latter makes use of some rather sophisticated results of algebraic topology, we shall obtain a simple proof of the existence of an equilibrium using a well-known lemma of -elementary combinatorial topology." This categorization of the mathematical register is surely of interest for subsequent curriculum development in  graduate courses in mathematical economics.} \cite{ga55}:  the first emphasized lower-semi-continuous correspondences and the second,   topological vector spaces, emphases that the chapter clearly imbibes.  With a quick recording of a fixed point theorem, and its subsequent application to the existence of a Nash equilibrium of a finite normal form game with payoffs generated by non-ordered preferences, it turns to showing   the non-emptiness of demand sets in consumer theory, and to an alternative proof of Geistdorfer-Florenzano's generalization of the Gale-Nikaido-Debreu (GND) lemma. The thrust of the chapter can best be captured in the (following) remark; see
     \cite[p. 21]{ya83a}. 

\bqu Although the relationship of open graph and openness of lower and upper section is known, the relationship of open sections with lower semi-continuity is still unknown. Below we examine this relationship.  \equ     
    
\nt     As far as mathematical economics  is concerned, rather than the results achieved,  the chapter is  pioneering in bringing the relevant tools to the subject, and its sensitivity to what then were regarded as purely technical (i.e. mathematical) subtleties.

However, it may be worthwhile for the reader to note, if only for the historical record,  that the  1983 chapter and the subsequent YP-Paper,  are largely tangential to the the preoccupation of the Ph.D itself, a dissertation of  six substantive chapters collected under the title  {\it Solution Concepts in Economic Theory.} Its 
explicitly-expressed gratitude to Professors Lionel W. McKenzie and George Metakides hints at its broad thrust: 

\bqu Professor McKenzie introduced me to General Equilibrium Theory and Professor Metakides to Logic and Nonstandard Analysis.\fn{Among others, Emmanuel M. Drandakis and S. A. Ozga are   two additional names that deserve mention for the record.   Yannelis writes,  \lq\lq I also wish to express my indebtedness to my first teacher of Mathematical Economics, Professor  Emmanuel M. Drandakis of the Athens School of Economics. My interest in General Equilibrium Theory was  inspired largely by my undergraduate with him.  Thanks are also due to the late Dr, S. A. Ozga of the London School of Economics, who as my supervisor, correctly advised me to come to Rochester." }  \equ 

\nt The acknowledgement is of interest in that it signals the application of mathematical logic, model theory in particular, to what Yannelis then saw as classical general equilibrium theory.\fn{We turn to Lionel McKenzie's influence on Yannelis in Section 4; the reader is also referred to Footnotes~\ref{fn:mck0},\ref{fn:mck1} and \ref{fn:mck2} below and the text they footnote.}  The chapter that we sight here  was  something for him to get out of the way before his substantive work of the role of a large number of agents in eliminating the convexity assumption in static general equilibrium theory, and in simultaneous-play normal form game theory, could begin. The essential subtext of the dissertation itself is {\it nonstandard analysis:} of the five remaining substantive chapters, three explicitly involve infinitesimal analysis, and the other two involve the Shapley-Folkman theorem.  Chapter III presents an existence  theorem for a \lq\lq nonstandard exchange economy" with a special subsection in the chapter on the \lq\lq novelty of a proof based entirely on Q-concepts,"  a passing consideration of consequence to subsequent history of this subject  that  one can  trace with some legitimacy to Anderson's  definitive paper.\fn{Yannelis  writes, \lq\lq Note that in contrast to Brown [11] and Khan [29], our proof is based entirely on Q-concepts. This gives us the opportunity to  appeal directly  to standard theorems using the \lq\lq transfer principle." \cite{an78} and in his subsequent {\it Handbook} survey articles explicitly credits Khan-Rashid.}      Chapter IV and V continue with a nonstandard exchange economy but with \lq\lq large" and \lq\lq small" traders, the first on \lq\lq non-discriminatory" allocations, and the second on \lq\lq fair allocations."\fn{We leave for future work carried out in the register of the  history of the application of nonstandard analysis to economics;  in particular, the specific advances Yannelis' dissertation makes over dissertations of a decade earlier, \cite{kh73}, \cite{ra76},   \cite{le77b},  \cite{ta78},   as well as those of his contemporaries    \cite{no84} and \cite{em85}, and follower \cite{ra98}.}      Chapters VI and VII turn to questions of  asymptotic implementation of the Shapley-value and Walrasian allocations, and answer them as   applications of the Shapley-Folkman-Starr theorem.

 With these remarks on portraiture, the chapter and the dissertation behind us,  we read  the YP-paper not  as a synecdoche  for Yannelis'  
  {\it oeuvre} even only upto the present-- the latter is a task  surely outside the authors' own competence, and perhaps outside the scope of a single essay.\fn{As we shall have occasion to emphasize in the last section of this  this paper, we neglect in particular all of Yannelis'  work on the economics of informations and incentives, on co-operative game theory and on decision theory.}  As such, the outline is simple: there are two parts to what we present: the theory and the applications. The highlights of the first are a framing of the 1983 paper in the light of current work and  two equivalence results spelling out  necessary and sufficient conditions for local selections that have so far proved elusive. The highlights of the second are four applications dictated by Yannelis' interests in classical Walrasian general equilibrium theory,  both in the dissertation and as they evolved in the corpus of his work so far: Berge's maximum theorem for utility functions {\it and} preferences, a generalization of the  GND  lemma  for discontinuous preferences,  an existence result for Shafer's take on the Walrasian setting, and an existence result for approximate  equilibrium  for a pure exchange economy with discontinuous preferences.


\section{The Mathematical Contribution} \lb{sec: theory}  The plan of this part of the work is straightforward: after a conceptual overview of the problem that revolves around Michael's selection theorem and considers a convex-valued lower semi-continuous correspondence defined on  a paracompact domain, we turn to an exegesis of the Yannelis-Prabhakar paper, track how it  feeds into work of four decades in both applied mathematics and mathematical economics, and in particular, extend his purview to show what extension it enables.

\subsection{The Conceptual Antecedents: Michael-Browder-Fan-Tarafdar}

This subsection revolves around four named theorems: a selection theorem, two fixed point theorems and an early result that inaugurated the application of the Hartman-Stampacchia  \lq\lq variational inequalities."   However, rather than the theorems themselves, our emphasis shall be on the underlying connections between the ideas that have been well-understood by the {\it cognoscenti} but have never been made explicit. In particular, we make explicit  the local-versus-global subtext, and   identify  neglected  contributions of Hakuhara and Tarafdar.

  But first things first: we begin with selection theorems and Himmelberg's  review of \cite{pa72},  in which  the context of  the problem that the author's book addresses is laid out. He writes: .  

\bqu Let $F:X \rightarrow {\cal A}(Y)$  be a function from a set $X$ to the collection of non-empty subsets ${\cal A}(Y)$ of $Y,$  i.e., for each $x \in X;\; F(x)$  is a non-empty subset of $Y.$ A function $f:X \rightarrow 
Y$  is called a selector for $F$ if $f(x) \in F(x)$ for all $x \in X.$ It is interesting and useful to know, given various continuity and measurability conditions on $F,$ that $F$ has selectors satisfying similar conditions. This book collects and organizes such theorems and provides examples of their application to stochastic games, and the solution of generalized differential  equations. \equ 

\nt He  justly describes Chapter 1 as a survey of the results of Michael on continuous selections for lower semi-continuous multi-functions   and singles out the following characterization of paracompactness in this chapter.   

\df
A Hausdorff space $X$ is {\bf paracompact} if each open covering of $X$ has an open, locally-finite refinement. 
\label{df: paracompact}
\dff

 \begin{customthm}{(Michael 1956)}[{}] 
Any Hausdorff  space $X$ is paracompact if and only if 
\ben[{\nf (i)}, topsep=0in]
  \setlength{\itemsep}{0in}
  \setlength{\parskip}{0pt}
  \setlength{\parsep}{0pt}

\item every open cover of $X$ has a  partition of unity subordinated to it, 
\ml every open cover of $X$ has a locally finite partition of unity subordinated to it, 
\item each lower semi-continuous function $X \rightarrow {\cal A}(Y )$, $Y$ a Banach space,  with closed convex values has a continuous selector.\fn{Himmelberg writes \lq\lq  measurable" instead of \lq\lq continuous; initially, we thought of keeping this slip in the statement of the theorem for  dramatic emphasis on how the measure-theoretic register  is intimately involved with the topological, and is always hovering in the background.  This is from Himmelberg's  review of the book in {\it Mathematical Reviews} available on MathSciNet;  in order to minimize references, we simply cite the review number, in this case MR0417368. Note that \cite{ss78} refer to a $T_1$-space as a Frechet space: we shall rely on  \cite{du66} and \cite{wi70} as our basic references for topological terminology throughout this work.}  
\een
\lb{thm: ro}
\end{customthm}

\noindent In the light of what we are seeing in this essay as the {\it leitmotiv} of Yannelis' contribution to mathematical economics and  to economic theory, this is a most fortuitous starting point, and brings out the intimate connection between paracompactness and the selection problem.  However, it is worth noting that Dieudonn\'{e}'s 1944 concept has proved elusive: a generalization of both compact and metrizable spaces, it is also intimately connected to the existence of {\it partitions of unity} and the extension problem. In terms of the way we are setting out the narrative here, what is of consequence is that it allows a move from the local to the global.\fn{As we shall see below, it is this aspect that will be focused on in \cite{hy16} and others. For a reader interested in paracompactness {\it per se,} we refer to \cite{mi04, mi10, mi11} and move on.}  Figure \ref{fig: paracompact} below illustrates commonly used topological properties that are weakening or strengthening on paracompactness.\fn{For the definitions of these topological concepts, see any of the standard texts sich as  \cite{du66}, \cite{mu00}, \cite{wi70} and \cite{ss78}. For the concept of stratifiability and its relations, see the work of Ceder and of Borges referenced in \cite{gr14}.} 

 \begin{figure}[htb]
\begin{center}
 \includegraphics[width=4in, height=2.2in]{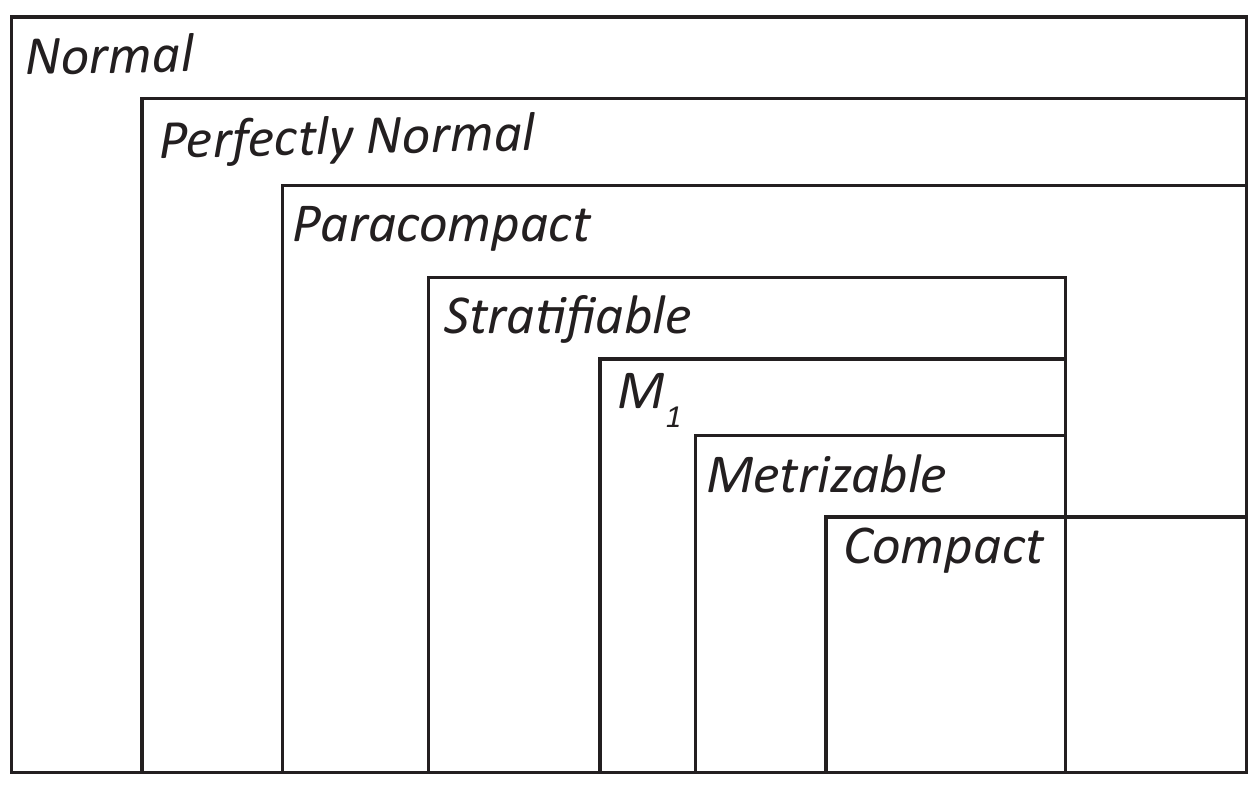}
\end{center}  
\vspace{-20pt}   
    
\caption{A Framing of Paracompactness}
\vspace{-5pt}   
   
  \lb{fig: paracompact}
\end{figure}

We now turn to a fixed point theorem that  serves as another  useful point from which to see the work that we report here: from  the  dissertation chapter  via YP-Paper  down to the present. Its  principal vernacular  --  selection, paracompactness, lower semi-continuity, infinite-dimensionality -- are all signature terms in the trajectory of this work.

 \begin{customthm}{(Browder 1968)}[{}] 
 Let $X$ be a non-empty, compact and convex subset of a Hausdorff topological vector space, and 
 $F: X\tra X$ a correspondence with non-empty and convex values such that  for each $y$ in $X$, $F^{-1}(y) = \{x\in X \ |\ y\in F(x)\}$ is open in $X$.  Then there exists $x^*\in K$ such that $x^*\in F(x^*)$. 
 \end{customthm}


A subtext of both of the results presented above is the postulate of continuity.   

\df 
Let $X, Y$ be non-empty  topological  spaces. A correspondence $P:X\tra Y$ 
\ben[{\nf (i)}, topsep=0in]
  \setlength{\itemsep}{0in}
  \setlength{\parskip}{0pt}
  \setlength{\parsep}{0pt}

\ml has the {\bf local intersection property} if for all $x\in X$ with $P(x)\neq \emptyset$ there exists $y^x\in Y$ and an open neighborhood $U^x$ of $x$  such that $y^x\in P(z)$ for all $z\in U^x$. 
  
\ml has {\bf open fibers} if for all $y\in Y,$ $\{x\in X|~ y\in P(x)\}$ is open. 

\ml has {\bf open graph} if for all $x\in X,$ $\{(x,y)\in X\times Y|~ y\in P(x)\}$ is open. 

\ml is {\bf upper semi-continuous (usc)} if $\{x\in X| P(x)\subseteq V\}$ is open for all $x\in X$ and all open $V \!\subseteq Y$. 

\ml is {\bf lower semi-continuous (lsc)} if $\{x\in X|  P(x)\cap V \! \neq \emptyset\}$~is~open for all $x\in X$ and all open $V \! \subseteq Y$. 
\een
\lb{df: continuity}
\dff

\nt It is by now well-known that the open graph property implies the open fibers property, which  implies the local intersection property.

\df
Let $X, Y$ be non-empty  topological  spaces. A correspondence $P:X\tra Y$ is {\bf Tarafdar-continuous} if there exists an open cover $\{O^y\}_{y\in Y}$ of $\{x\in X|P(x)\neq\emptyset\}$ such that $O^y \subseteq P^{-1}(y) = \{x\in X | y\in P(x)\}$ for all $y\in Y$. 
\lb{df: tarafdar}
\dff

Next, we show that Tarafdar-continuity and the local intersection property are equivalent. 
\prp
Let $X, Y$ be non-empty topological  spaces and $P: X\tra Y$ a  correspondence. Then $P$ is Tarafdar-continuous if and only if it has the local intersection property. 
\lb{thm: tarafdarc}
\prpp

\noindent  This result highlights that the local intersection property was first introduced in \cite{ta77}. This property also goes by other names in the literature; a terminological cleaning up is overdue.\fn{\cite{tz95} refer to it as the    {\it transfer open lower sections} and   \cite{uy16a} calls it the  {\it constant neighborhood selection property}; see also  \cite{mt87} and \cite{ta91}. It is worth noting that \cite{mt87} provide equivalence of various theorems that use versions of local intersection property, but they do not explicitly show the equivalence of the continuity assumptions without the presence of other assumptions on the correspondence and the spaces. The discussion and example above highlight the importance of such consideration.}   

\prf[Proof of Proposition \ref{thm: tarafdarc}]
Assume $P$ is Tarafdar-continuous. Then there exists an open covering $\{O^y\}_{y\in Y}$ of $\{x\in X|P(x)\neq\emptyset\}$  such that $O^y \subseteq P^{-1}(y) = \{x\in X | y\in P(x)\}$ for all $y\in Y$. Pick $x\in X$ such that $P(x)\neq \emptyset$. Since there exists some $y\in Y$ such that $x\in O^y$, and $y\in P(z)$ for all $z\in O^y$,   $P$ has the local intersection property. 

Now assume $P$ has the local intersection property. Then for all  $x$ with $P(x)\neq\emptyset\}$, there exists $y^x\in Y$ and an open neighborhood $U^x$ of $x$  such that $y^x\in P(z)$ for all $z\in U^x$. Define $\hat Y$ as the collection of all such $y^x$, i.e., $\hat Y=\{y^x| x\in X, P(x)\neq\emptyset\}$. For any $y\in \hat Y$, let $X^y=\{x\in X|y^x=y\}$.  Define $O^{y}=\bigcup_{x\in X^y} U^x$ for all $y^x\in \hat Y$ and $O^y=\emptyset$ for all $y\notin \hat Y$. It is clear that $\{O^y\}_{y\in Y}$ is an open covering of $\{x\in X|P(x)\neq\emptyset\}$  and $O^y \subseteq P^{-1}(y) = \{x\in X | y\in P(x)\}$ for all $y\in Y$. Therefore, $P$ is Tarafdar-continuous. 
\prff

Tarafdar-continuity, as expressed above, is equivalent to the version stated in \cite{ta77} when the correspondence $P$ has  non-empty values, which is assumed  in Tarafdar's theorem, and indeed in most fixed point and selection theorems. 
For correspondences with empty values, the version we present above is weaker.   That is, if we state the definition by replacing the set $\{x\in X|P(x)\neq\emptyset\}$  with $X$, then Tarafdar-continuity will be stronger than the local intersection property for correspondences which have empty values. In this case, the proof of the proposition above shows that local intersection property is implied by Tarafdar-continuity. However, the converse relationship does not  hold. In order to see this, let $[0,1]$ be the unit interval with the usual topological structure and the correspondence $P:[0,1]\tra [0,1]$ defined as $P(x)=1$ if $x>0$ and $P(0)=\emptyset$. It is clear that $P$ has open fibers, hence has the local intersection property. However, for all $y\neq 1$, $O^y=\emptyset$ has to hold and $O^1$ can be at most $(0,1]$. Hence $\{O^y\}_{y\in [0,1]}$ cannot yield an open covering of $[0,1]$. Therefore, $P$ is not Tarafdar-continuous.

We can now present the fixed point theorem that motivated Tarafdar-continuity.\fn{For Tarafdar's perspective on fixed point theory, see \cite{tc08}. Even though \cite{bo85} is outdated, the reader may find his Chapter 8 on variational inequalities useful, though the reader should bear in mind \cite{ya85} is not in the book, and even though he cites his YP-Paper, he does not give it quite the importance that it deserves.} 

\medskip

\nt {\bf Theorem} ({\bf Tarafdar 1977}). {\it Let $X$ be a non-empty, compact and convex subset of a Hausdorff topological vector space. Let $P: X\tra X$ be a Tarafdar-continuous correspondence with non-empty and convex values.  Then there exists $x^*\in K$ such that $x^*\in P(x^*)$.} 

\medskip

  So far we have presented as our mathematical antecedents, a selection theorem due to Michael, and two fixed point theorems, one attributable to Browder and the other to Tarafdar.  However, as already noted, Yannelis  relied on Fan, and we turn to a 1969 result of his that has some consequence for the narrative that we are in the process of developing.   

\df[\bf Browder 1954] Let $X$ be a reflexive complex Banach space and $X^*$ its dual. A function $T: X\ra X^*$ is {\bf demi-continuous}  if it is continuous from the strong topology of $X$ to the weak topology of $X^*$.
\dff

\df[\bf Fan 1969]
Let $E$ be a real Hausdorff topological vector space, and let $X \subseteq E$. A correspondence $F: X\tra E$ with non-empty values is {\bf upper demi-continuous} if for every $x\in X$ and any open half-space\fn{Note that an open half-space $H$ in $E$ is a set of the form $H=\{x\in E| \psi(x)> r\}$, where $\psi$ is a continuous linear form on $E$, not identically zero, and $r$ is a real number. For details consider \cite{tc08}.} $H$ in $E$ containing $F(x)$, there is a neighborhood $N$ of $x$ in $X$ such that $F(x')\subseteq H$ for all $x'\in N$. 
\dff

\nt   It is not difficult to see that an usc correspondence is upper demi-continuous. However, an upper demi-continuous correspondence need not be usc; see \citet[Example 17.39, p. 575]{ab06}.  The relationship\fn{For the definition of {\it quasi upper semi-continuity}, see \cite{po97}. The idea is the same as {\it Hausdorff upper semi-continuity} presented in \cite{su97}. This concepts goes back to \cite{ae84}, see \cite{ya90} for details. Moreover, for a comprehensive discussion of the relationship between different  continuity postulates in economic theory, see \cite{ku19b}.} between these concepts is  illustrated in Figure \ref{fig: continuity}.

 \begin{figure}[htb]
\begin{center}
 \includegraphics[width=4.8in, height=2.2in]{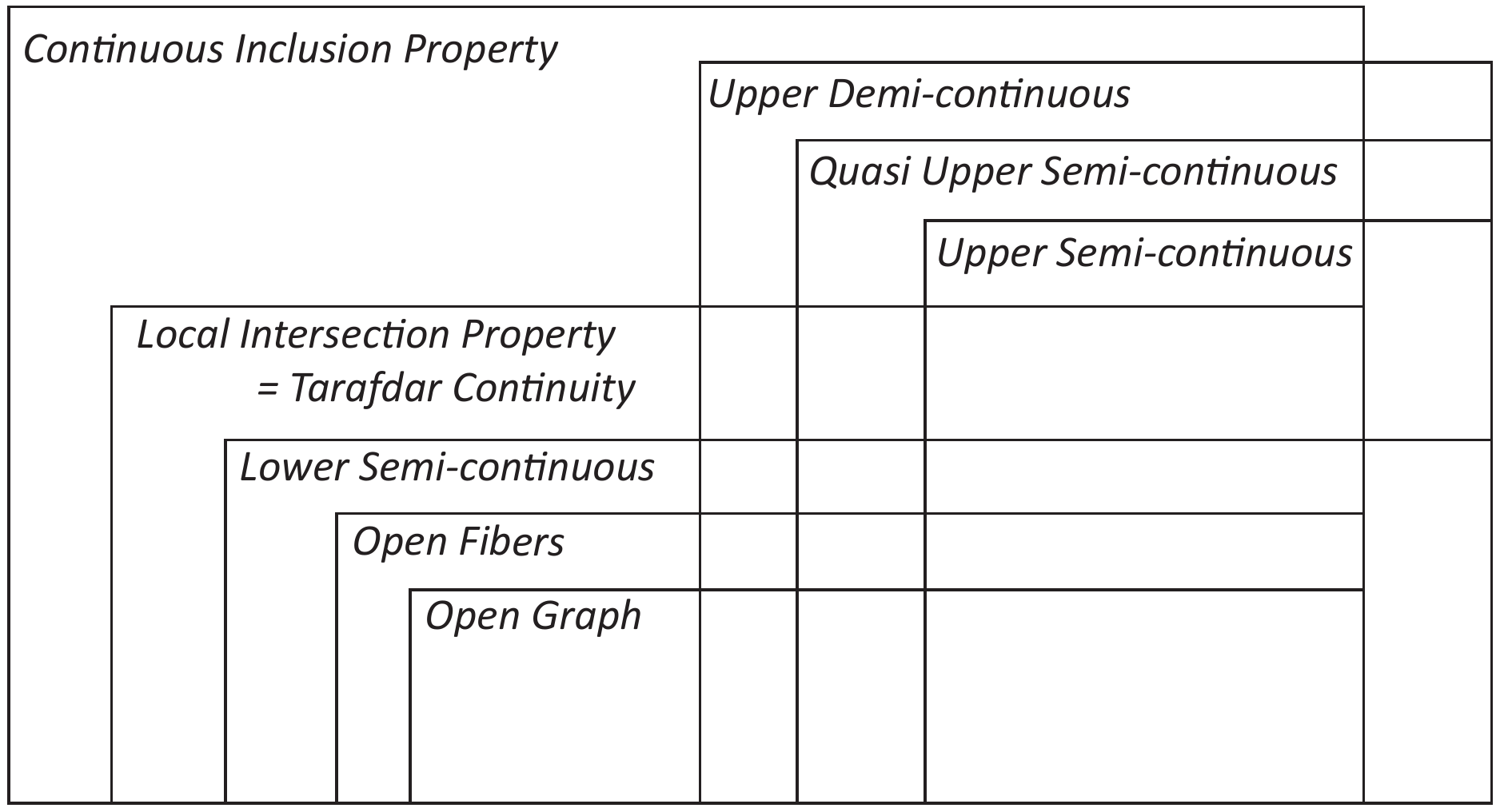}
\end{center}  
\vspace{-20pt}   
    
\caption{Continuity Concepts for Correspondences}
\vspace{-5pt}   
   
  \lb{fig: continuity}
\end{figure}

It is conventional wisdom in the profession that for questions pertaining to existence of solution concepts, fixed point theorems constitute the  first line of attack, and  the geometric form  of the Hahn-Banach theorems typically  reserved for their price characterizations.  An early result 1969 result of   Ky Fan's  diffused such a sharp 
line,\fn{As we shall see below, \cite{mc81a} and  \cite{ya85} were the firsts to breach this line in their proofs of the GND lemma. Mas-Colell's proof of the existence of the bargaining set furnishes another example.  } and brought the separation theorem to bear on the fixed point theorem, and in doing so, became  a direct precursor of the application of the Hartman-Stampacchia variational
 inequalties.\fn{We refer the reader for a comprehensive discussion of this subject and its application to mathematical economics to    \cite{hs66}, \cite{tc08}, and  {\cite{mc20} for a state of the art treatment.} We shall return to this subject in Section 3.2 below.}   But before the result, the question to which it is an answer; the  problem that furnishes the 
 context.\fn{As \cite{mc20} writes \lq\lq The Ky Fan equilibrium problem (aka the Ky Fan Inequality) is a fundamental result of non-linear analysis with myriad applications in optimization theory, fixed point theory, mathematical economics and game theory to name just a few."  Also see Footnote~\ref{fn:rich1} below, and the text that it footnotes.
 \label{fn:rich0}   }
 
 \medskip
 
\noindent {\bf Ky Fan Equilibrium Problem:} Given a set $X$ and a function $f\!:\!X\!\times \!X\ra \Re$, find $\bar x \in X$ such that $f(\bar x, x)\leq 0$ for each $x\in X$.

\medskip 

This problem can be seen in a slightly general setting. 

 \medskip
 
\noindent {\bf A Variational Inequality:} Given two sets $X$ and $Y$ and a function $f\!: \!X\!\times\! Y\ra \Re$, find $\bar x \in X$ such that $f(\bar x, y)\leq 0$ for each $y\in Y$.

\medskip 

However, we can now consider a concretization,

\medskip
 
\noindent {\bf A Variational Inequality for TVS:} Let $X$ be a locally convex topological vector space with $X^*$ its dual space, $K \subseteq X$ a non-empty compact convex set, a continuous mapping $T: K \rightarrow X^*$  find $\bar x \in X$ such that $\langle T(\bar x,), \bar x - y \rangle\leq 0$ for each $x\in X$.

\medskip 

\noindent  \cite{br67} offers a generalization of the fixed point theorems of Schauder and Tychonoff and its reviewer  writes that they are consequences of a theorem that solves the above problem, and that the proof of  this Browder's theorem  \lq\lq uses a continuous partition of unity in a fashion reminiscent of M. Hukuhara's elegant proof of the Schauder theorem."\fn{For Belluce's review, see MR223944 of Browder's paper.     \label{fn:rich1}} 

We can now present 

 \begin{customthm}{(Fan 1969)}[{}]
 Let $X$ be a non-empty compact convex set in a Hausdorff topological vector space $E$. Let $F, G$ be two upper demi-continuous set-valued mappings defined on $X$ such that:
\ben[{\nf (i)}, topsep=0in]
  \setlength{\itemsep}{0in}
\ml For each $x \in X$, $F(x)$ and $G(x)$ are non-empty subsets of $E$.
\ml For every $x\in X$, there exist three points $y\in X$, $u\in F(x), v\in G(x)$ and a real number $\lambda >0$ such that $y- x=\lambda(u- v)$.
\een
Then there exists a point $\bar x\in X$ for which $F(\bar x)$ and $G(\bar x)$  cannot be strictly separated by a closed hyperpIane, i.e., we cannot find a continuous linear form $\psi$ on $E$ and a real number $r$ such that $\psi(x)<r$ for $x\in F(\bar x)$ and $\psi(y)>r$ for $y\in G(\bar x)$.

\end{customthm}

\begin{wrapfigure}{r}{6.7cm}

\vspace{-25pt}

\caption{Example for Fan's Theorem}
  \lb{fig: demicontinuity}
~~~\includegraphics[width=6.1cm, height=5cm]{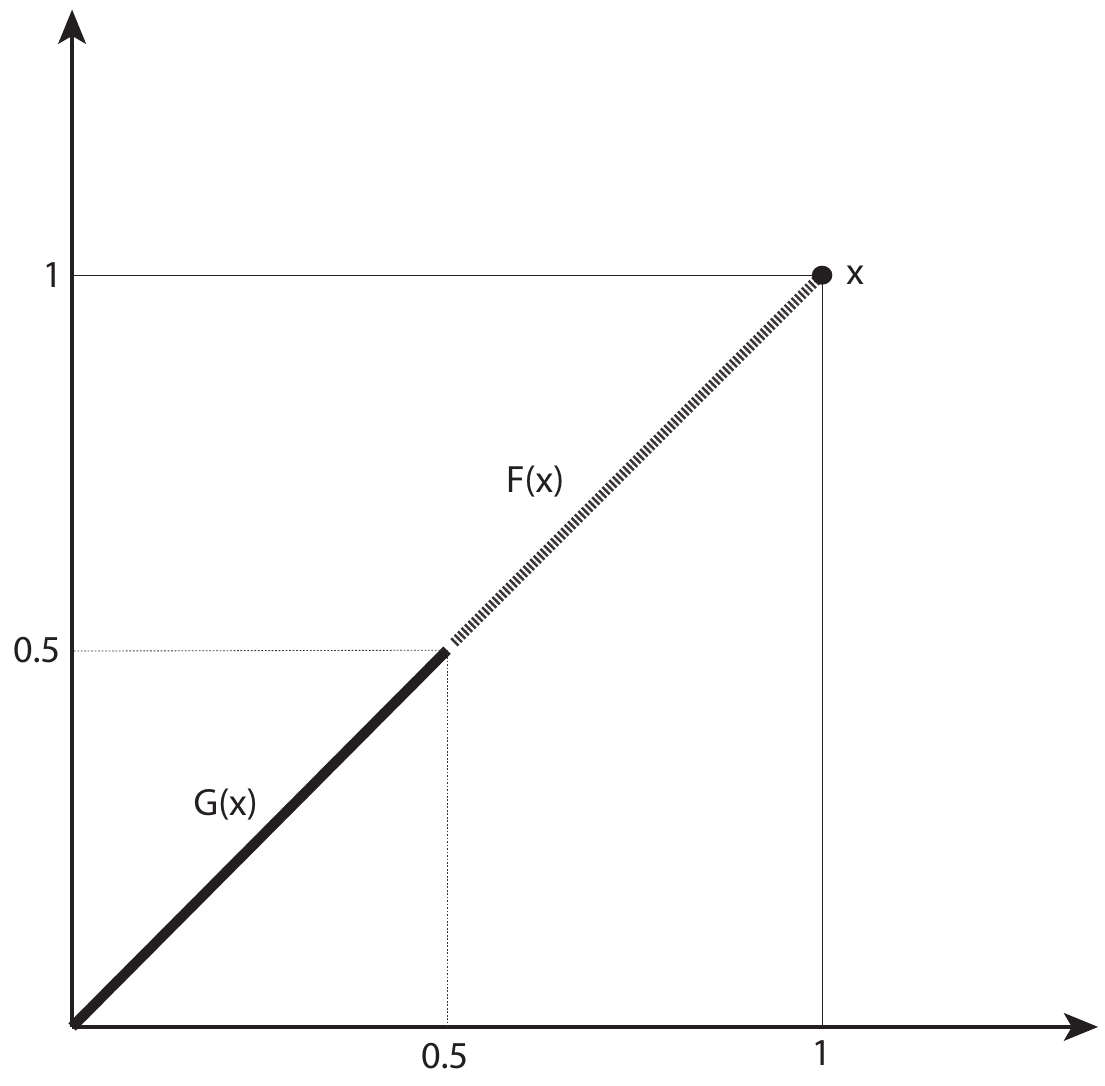}
\lb{fig: demic}
\end{wrapfigure}

 Readers may find the following example useful in understanding  assumption (ii) in Fan's theorem.  Let $E=\Re^2$ with the usual linear and topological properties and $X=[0,1]^2$. Define correspondences $F, G$ as $F(x)=\{y\in E|0.5<y_1=y_2\leq 1\}$ and $G(x)=\{y\in E|0\leq y_1=y_2\leq 0.5\}$ for all $x\in X$. Pick $x=(1,1)$. For any $u\in F(x), v\in G(x)$ and $\lambda>0$, $\lambda (u-v)=(a,a), a\in (0,0.5)$. It is easy to see that for all $y\in X$, $(y-x)_i\leq 0$ for some index $i=1,2$. Therefore, assumption (ii) fails. Note also that $F(x)$ and $G(x)$ cannot be strictly separated by a closed hyperpIane for all $x\in X$.    
As you see in this example, as illustrated in Figure \ref{fig: demic}, the algebraic boundary of $X$ plays a crucial role. 
In fact, for any other point in $X$, assumption (ii) is automatically satisfied; see \citet[p. 237]{fa69}.\fn{We return to demi-continuity in Section 3.2 on the generalizations of the Gale-Nikaido-Debreu Lemma.} 

%
%

\subsection{The Initiation:  Yannelis-Prabhakar (1983)}   

Thanks to \cite{ma74},   non-ordered preferences were very much in the neoclassical air in the decade since 1974, and with this as our point of departure,  we begin the exegesis of  YP-Paper   with the contemporaneous dissertation  chapter\fn{In the sequel, we shall simply refer to YP-Paper as  the \lq\lq paper," and to Chapter 2 of Yannelis' dissertation as the \lq\lq chapter."} 
   is solely 
in the register of Walrasian equilibrium theory with a  \lq\lq weak form of Walras' law":    the commonality lies in his invocation and use of Fan's generalization, as in the  chapter.\fn{For a detailed reference and discussion on weak and strong versions of Walras' Law see  \cite{mr08} and \cite{co20}, and see also  \cite{mc81a}, \cite{ya85}, \cite{mt87}, \cite{uy16a} and \cite{hy17}.} 

We now turn to the {\it pi\'ece de resistance} of this tribute, and single out the following result in YP-Paper.  To the authors knowledge, this was the first time that the notion of a {\it paracompact} set was used in mathematical economics.

 \begin{customthm}{(Yannelis-Prabhakar 1983)}Let $X$ be a paracompact  Hausdorff space and $Y$ be a topological vector space. Suppose $P:X\tra Y$ is a non-empty and convex valued correspondence with open fibers. Then $P$ has a continuous selector. 
 \end{customthm}

 \noindent Presents the proof. 
(1) The Brouwer fixed point theorem, in its classical single-valued form.
(2) The existence for a finite covering of a compact space of a partition of
unity subordinated to this covering.   Indeed, one can sight Michael and Browder    as the two mathematicians.  As it happened there was an additional set of lecture notes published two years earlier in which Michael has his own overview of the problem.

The composition of the paper is straightforward: titled \lq\lq  Existence of maximal elements and equilibria in linear topological spaces," it departs from the chapter in a  move beyond   (i) the finiteness assumption and in allowing  an infinite number of commodities and a countably infinite number of agents, (ii) Walrasian equilibrium theory to the abstract economy setting of \cite{de52} and \cite{ss75}, and (iii) Fan and Michael to an explicit invocation of   the  \cite{br68} and his  open-fiber assumption as well as that of lower semi-continuity of a correspondence.  
We say that the correspondence  $E: D \rightarrow \rightarrow \R^n$   is upper demi-continuous if for each open half-space $B$  in $H \in \R^n,\;
E^+[H] =  {p : E(p) \subseteq E}$ is open in $D,$" a definition
  due to \cite{br64} and \cite{fa69}.\fn{For closed-valued correspondences this definition is
equivalent to what Geistdoerfer-Florenzano calls {\it upper hemicontinuity.}" -- we'll have more to say on this in the next section. \label{fn:mf}  }

  In terms of a more detailed listing, after the preliminary first and second sections, the chapters' echo is clearest in Sections 3 and 4 that are devoted to  selection and  fixed point theorems: a lot of the definitions are simply transcribed directly from the chapter.  However, it is in the elimination of the  the section on the GND lemma, and transmutation of a {\it game} into an {\it abstract economy} that Sections 5 and 6 constitute the {\it pi\`{e}ce de resistance} of the paper. Section 7 lists examples of topological vector spaces, and is more a reflection of the professional unfamiliarity   with functional analysis in the infinite-dimensional mode, even  eleven years after Bewley's 1972 paper.

In terms of a bald statement, the Yannelis-Prabhakar paper is pioneering\fn{
Given the importance that the authors are attaching attach to the YP-Paper, they feel it is incumbent on them to be crystal-clear about how it relates to the antecedent literature, and in particular to \cite{br68}; they invite the reader to compare the proof of Theorem 3.3 in  the YP-Paper and  the proof of Theorem 1 in \citet{br68}. The reader should also note Footnote 4 in the YP-Paper for its reference to Browder's work.}   
on several counts: (i) its invocation of the lower semi-continuity of a preference correspondence, (ii) in its systematic use of Michael's selection theorem, (iii)  in its assumption  of the paracompactness of an action set, and (iv) in its topological sensitivity  in the dropping of the   a locally-convex adjective, and considering only topological vector space. 
But all these are micro--criteria; at the macro level, the biggest contribution of the YP-Paper and the Y-Chapter is its resolute move from finite to infinite-dimensions, one that was to lead only two years later in \cite{ya85} an infinite-dimensional version of the GND.\fn{See Footnote \ref{fn:fin1} for further elaboration.        \label{fn:fin0}}

\subsection{The State of the Art: He-Yannelis (2016) and Cornet (2020)}   \label{sec: state}
Our focus in this subsection is to identify the traces of the 1983 work in  the statement and understanding of the subject as of now:  what is it in current and ongoing work   that would be new to the 1983 authors, as in both the Y-Chapter and the YP-Paper as far as the selection theorems and their applications are concerned?  We begin with the property encapsulated in the following definition  due to \cite{hy16}. 
\medskip

\df
Let $X, Y$ be non-empty subsets of a topological vector space and $P:X\tra Y$ a correspondence. The correspondence $P$  has the {\bf continuous inclusion property} 
if for all $x\in X$ such that $P(x)\neq \emptyset,$ there exists an open neighborhood $U^x$ of $x$ and an usc correspondence $F^x:X\tra Y$ which has non-empty, convex and closed values and $F^x(y)\subseteq P(y)$ for all $y\in U^x.$
\dff
 
 \noindent We provide five different continuity postulates on correspondences in Definition \ref{df: continuity}.  For topological vector spaces,  any of the five conditions above, in the presence of some other assumptions which are standard in fixed point and selection theorems, implies the continuous inclusion property; see  \citet{hy16} for details.\footnote{Analogous versions of this  property have been assigned different names: {\it correspondence security} in \citet{bs09, bm13},  {\it continuous security}  in \citet{re16a},  \citet{cp16} and {\it continuous neighborhood selection property}  in \cite{uy16a}.  The last work provides a fixed point theorem based on this weak continuity assumption; see \citet[Definition 22 and Theorem 9, pp. 93--94]{uy16a}.  
}   
 The reader should note well how the notion of a local selection {\it function} from a correspondence has been naturally uplifted to the idea of obtaining a local selection correspondence from a correspondence. Now with this this notion at hand, He-Yannelis unify the fixed point theorems of Browder and Kakutani-Fan-Gliksberg for locally convex Hausdorff  topological vector spaces in the following way.
 
 \smallskip
 
 \begin{customthm}{(He-Yannelis 2017)}[{}] 
Let $K$ be a non-empty, convex and compact subset of a Hausdorff locally convex   topological vector space and $T:X\tra X$ a correspondence with non-empty, convex values and have the continuous inclusion property. Then there exists $x^*\in K$ such that $x^*\in T(x^*).$    \end{customthm}

As is well-understood, the Kakutani-Fan-Glicksberg fixed point theorem assumes the relevant mapping to be usc, whereas the  Browder theorem    assumes 
that the mapping  has open fibers, and the above theorem is a unification simply because the   continuous inclusion property is weaker than either of these assumptions. Neither assumption implies the other. 
  In order to see that upper semi-continuity does not imply the  open fiber condition,    consider the following example. Let $[0,1]$ has the usual topology and $P:[0,1]\tra [0,1]$ is defined as $P(x)=\{x\}$ for all $x\in [0,1].$ It is clear that $P$ is usc and it does not have open fibers. In order to see that the open fiber condition does not imply upper semi-continuity,  consider the following example. Let $[0,1]$ has the usual topology and $Q:[0,1]\tra [0,1]$ is defined as $Q(x)=\{x'|x'>x\}\cup \{1\}$ for all $x\in [0,1].$   It is clear that $Q$ has open fibers and is not usc. 

The continuous inclusion property assumes that the relevant correspondence has `nice' local selections. As mentioned  in Section 2.2,   a special form of local selections,  which assumes the local selections are constant, was first presented by \cite{ta77} in the context of  a fixed point theorem and variational inequalities. Here we are focussing  on the post-Tarafdar period; see for example \cite{mt87} and  \cite{ws96} for  fixed point theorems that are based on constant selections and \cite{pa99}  for fixed point theorems that are based on single-valued continuous selections. The fixed point theorem above  not only generalizes and unifies the fixed point theorems of Browder and Kakutani-Fan-Glicksberg but also  those of furnished in Tarafdar, Wu-Shen and Park.\fn{The content of the last two paragraphs is simply  essentially  copied from  \citet[pp. 95--96]{uy16a}} 

\medskip

\citet{br68} proved his theorem for Hausdorff topological vector spaces. The theorem above is a generalization of Browder's theorem for locally convex Hausdorff topological vector spaces. 
A natural question arises at this stage whether the He-Yannelis result is valid for  topological vector spaces that are not necessarily locally convex or Hausdorff. In the context that Browder's work, it is a source of satisfaction that \citet[Theorem 1]{ta77} showed that the following  full generalization is possible. 
 
\vspace{5pt}

\nt {\bf Theorem} ({\bf Tarafdar 1977}).  {\it 
Let $X$ be a non-empty, convex and compact subset of a Hausdorff topological vector space and $P:X\tra X$ a correspondence with non-empty, convex values and have the local intersection property. Then there exists $x^*\in X$ such that $x^*\in P(x^*).$} 
\medskip

\noindent Note that Tarafdar's theorem pertains to Hausdorff topological vector spaces, and \citet[Theorem 6]{bm05} and \citet[Theorem 10]{uy16a} generalized it to show that  the Hausdorff  separation axiom is redundant.\fn{\cite{ta77} generalized Browder's theorem by assuming correspondences satisfies a continuity property which we show in Proposition \ref{thm: tarafdarc}  is  equivalent to the local intersection property. Also, \cite{ws96} proved a version of this  theorem  for locally convex Hausdorff topological vector spaces; see also \citet[Section 4.7]{tc08} and \cite{pr11} for recent treatments. The authors thank Rich McLean for pointing out  \citet{bm05} after the first draft of this manuscript was circulated. We invite the reader to check that the methods of proof of Balaj-Muresan and Uyanik embody different ideas.}   
 
 Next, we turn to  the antecedent  literature. The continuity postulate is the standard technical assumption in order to obtain existence of an equilibrium. Earlier work assumes the (weak) preference relation is complete transitive and has closed sections, which is equivalent to closed graph assumption, and also to the openness of the graph of the asymmetric part of the weak relation. However, when completeness and/or transitivity is relaxed, one needs to be careful as to which continuity assumption is to be used.\footnote{On the one hand they are not equivalent, on the other hand assuming too much continuity may imply completeness and transitivity; see \cite{ku19a}.}  For non-ordered preferences,  \citet{fa61} uses open graph assumption in order to obtain a maximal element in a tvs. \citet{so71} replaces it with the open fibers assumption. \citet{ma74}, \citet{gm75} and \citet{ss75} use open graph assumption in order to obtain the existence of an equilibrium in games and economies. Y-Chapter and YP-Paper replace it with open fibers for topological vector spaces and with lower semi-continuity for finite dimensional topological vector spaces. See \citet{be92} for an exposition.  
Finally, there is an alternative literature started with \cite{bk76} and \citet{to84}, where the traces can be found in the proof of the main result of \cite{gm75}, moving in the opposite direction and obtaining a {\it majorization}, a ``nice" covering relation. In this branch of the literature, any maximal element under the new covering relation is required to be a maximal element under the original preference relation. This is {\it not} the direction we pursue in this paper. 
The following table provides a portrait of the selection theorems\footnote{See also \citet{ss75, ss76, ss82}, \cite{ta77, ta91}, \cite{mt87}, \citet{uy16a}, \citet{mo75, mo07}, \citet{mc02}, \citet{btz93}, \citet{ss75}, \citet{ge86}, \cite{tz92}; \cite{pa99, pa11}, \cite{dp02},  \cite{ag06} and \cite{sc15b} for results on continuous selections and existence of an equilibrium in games and economies. } and their applications. Its objective is not to discuss the priority of the results but simply to connect.\fn{See Terry Tao's 2019 survey, \citet*[Figure 1]{detal19}, of a basic identity in linear algebra for an example of result in mathematics which has been proved several times independently.}

\begin{table}[ht]
\begin{center}
\begin{tabular}{l|l} 
Continuity Assumption & ~~~~~~~~ Continuous Function \\ 
\hline \hline 
Open Graph & Gale-Mas-Colell (1975),  Yannelis (1983) \\
Open Fibers & Browder (1968),  Yannelis-Prabhakar (1983) \\
Lower Semi-continous  & Michael (1956),  Yannelis-Prabhakar (1983) \\
Constant Local Selections  & Tarafdar (1977),  Mehta-Tarafdar (1987)   \\
  & Tian-Zhou (1995), Wu-Shen (1996), Uyanik (2016)   \\
Continuous Local Selections  & Park (1999) \\
\hline  

& ~~~~~~~~ Upper Semi-Continuous Correspondence \\
\hline \hline 

Upper Semi-Continuous   & Barelli-Meneghel (2013), He-Yannelis (2016, 2017),  \\
Local Selections  & Carmona-Podczeck (2016),  Reny (2016),   \\
  & Uyanik (2016), McLean (2020), Cornet (2020)  \\

\hline \hline  
\end{tabular}
\end{center}
\vspace{-10pt}

\caption{Continuous Selection Theorems}
\lb{tbl: relation}
\end{table}  
  

In the remaining part of this subsection we engage with the definitive analysis of \cite{co20} on the Gale-Nikaido-Debreu lemma and frame it in the light of the antecedent work of \cite{mc81a}, \cite{ya85}, \cite{mt87}, \cite{uy16a} and \cite{hy17}. We do it not only for its own intrinsic interest but also as a  preparatory step  for the generalization of the Gale-Nikaido-Debreu lemma presented here and in Section 3.2. We begin with a prescient statement made more than 25 years ago in a consideration of Lionel McKenzie's work on the existence of Walrasian equilibrium. 

 \citet[pp. 34--35]{kh93} is worth quoting at some length: 
	
\bqu The conjunction of Brouwer's fixed point theorem and the separating hyperplane theorem that is at the heart of McKenzie's 1959 proof raises the question whether his technique can be used to give an alternative proof of the (so-called) Debreu-Gale-Nikaido lemma. Such a proof has been offered in several recent contributions. The basic idea is simple.\fn{Khan continues, \lq\lq  If   the excess demand set does not intersect with the negative orthant at any price, the two sets, being closed and convex, can be strictly separated. Consider the set of these normalized separating hyperplanes, one set for each price we began with. This correspondence has a continuous selection and hence a fixed point. But this furnishes a contradiction to Walras' law. The point of contact with McKenzie's 1959 proof is that he, too, is selecting continuously a unique hyperplane and this selection is obtained, as in McCabe, by construction. Mehta and Tarafdar use the more powerful Tarafdar fixed point theorem, and Yannelis offers, and relies on, a more sophisticated selection theorem." The reader can pick up this discussion in Section 4 below.}  \equ  

\citet[p.2]{co20}  provides a state-of-the-art  reading of the literature on the  Gale-Nikaido-Debreu  lemma; he notes that the   classical proofs of this lemma use either a fixed point theorem, or K-K-M lemma; see for example \cite{de82, ge82, fl86, ky94}. \citet{mc81a} and \cite{ya85} provided a simple alternative proof by using the separating hyperplane theorem and a fixed point theorem. They   introduced a `nice' mapping {\it Browder-McCabe-Yannelis map},\fn{The Browder-McCabe-Yannelis map is related to variational inequalities and the traces can be found in \cite{br67, br68} which relies on Hakuhara; see Footnote~\ref{fn:rich0}.  In particular, the mapping is given as the $N$ correspondence in \citet[p. 287]{br68} and in \citet[p. 287]{br67}, as the $\Psi$ correspondence in \citet[p. 168]{mc81a} and  as  the $F$ correspondence in \citet[p. 598]{ya85}. All obtain a continuous selection of the relevant correspondence by appealing to partition of unity, either directly or indirectly,   to obtain a fixed point.  \label{fn:rich2}} which assigns for each price $p,$ the set of prices at which the value of the excess demand is strictly greater than zero. It also allows them to use a weaker verison of Walra's law. \cite{mc81a} assumed the excess demand correspondence is usc which implies the Browder-McCabe-Yannelis map has open fibers. Later,  \cite{mt87} extended  \cite{ya85} by directly imposing a weaker continuity assumption on the Browder-McCabe-Yannelis map, rather than the excess demand correspondence.

In  addition to the GND lemma, Cornet writes an equivalence result that shows that the generalization of the GND lemma is {\it equivalent}  to Kakutani's fixed point theorem. He wrote ``The proof of Theorem 1 relies on Kakutani's theorem and is in fact equivalent to it, as shown in Remark 2."\fn{For an elaboration of this interesting contribution to the mathematics literature on fixed point theory, see \citet[pp. 2 and 11]{co20} and \citet[Section 2.1]{ma09}.} We now turn to his principle result.

 \begin{customthm}{(Cornet-2020)}[{}] 
 Let $P\subseteq \Re^\ell$ be a nondegenerate,  closed, convex, cone, with vertex $0$. Let $Z$ be a correspondence, from \text{\nf co}$[P \cap S]$ to $\Re^\ell$, with non-empty, convex, compact values, satisfying the continuous inclusion property and Walras' law on $P\cap S$. Then there exists $p^*\in  \text{\nf co}[P \cap S]$ such that $Z(p^*) \cap P^0 \neq \emptyset$.  \end{customthm}

\nt As \cite{co20} points out, he limits himself to finite dimensional spaces; we remind the reader that \cite{ya85} and \cite{hy17} are infinite dimensional.\fn{For this constant drum beat on infinite-dimensionality, see Footnotes \ref{fn:fin0} and \ref{fn:fin1} and the text they footnote.} 
 Cornet's ideas and proofs are novel, an elaboration on this is needed; especially his Claim 2 and Lemma 2. Cornet pioneering steps follow the path originally laid out in \cite{mc54} and delineated in \cite{kh93}. These steps are of profound consequences for the extension of the Walrasian general equilibrium theory to increasing returns to scale production technologies in which Cornet is a pioneering scholar with significant contributions.\fn{See \cite{kh87}, \cite{co88},  \cite{vo92}  and their references.}  In his Remark 1, Cornet identifies the joint continuity of a function that maps price and excess demand to the real line. He mentions that this property ``may not hold in infinite dimensional spaces, hence the importance of considering the weak version of Walras' law" which we return in Section 3.2 below.    
  What the reader should appreciate that Cornet is also not working on the price simplex but also, as in Debreu (1956) on a cone.\fn{\citet[p. 2]{co20} wrote ``[...]  we allow for the cone $P$ of admissible prices to be a non-degenerate closed, convex, cone of vertex 0 in the finite dimensional commodity space $\Re^\ell$." He also essentially reproduced this sentence in his abstract. }  
 
 Following the pioneering work of Browder-Fan, \cite{ya85} uses demi-continuity in the generalization of the GND lemma to infinite dimensional commodity spaces. We reproduce the result for the reader's convenience.

 \begin{customthm}{(Yannelis-1985)}[{}] 
 Let $X$ be a Hausdorff locally convex linear topological space, $C$ a closed, convex cone of $X$ having an interior point $e$, $C^* =\{p\in X^*~: ~p\cdot x<0 \text{ for all } x\in C\}\neq \{0\}$ the dual cone of $C$ and $A= \{p\in C^*~:~ p\cdot e = -1\}$. Let $\zeta: \Delta \ra 2^X$ be a correspondence such that:
\ben[{\nf (i)}, topsep=0in]
  \setlength{\itemsep}{0in}
  \setlength{\parskip}{0pt}
  \setlength{\parsep}{0pt}

\ml $\zeta$ is upper demi-continuous in the weak$^*$ topology, 
\ml $\zeta(p)$ is convex non-empty and compact for all $p\in \Delta$, 
\ml for all $p\in \Delta$ there exists $x\in \zeta(p)$   such that $p\cdot x\leq 0$.
\een
Then there exists $p^*\in \Delta$ such that $\zeta(p^*)\cap C\neq \emptyset.$
\lb{thm: GNDYannelis}
\end{customthm}
 
\nt \cite{po97}  comments on this theorem as follows.  

\bqu  
 Note that Theorem 3.1 in Yannelis (1985) only requires $\langle p, z\rangle\leq 0$ to hold for some $z\in \zeta(p)$ but not for all  $z\in \zeta(p)$. As shown by our constructions, this allows to replace demand correspondences by objects with enough continuity properties. In this respect this theorem offers, in particular, a powerful tool to overcome the problem that the wealth map may not be jointly continuous in an infinite dimensional setting.\fn{We do not want to be overly defensive about this: as emphasized by Halmos and others it is this dialectical interplay between the finite and the infinite that leads to a synthetic development for the subject. \label{fn:fin1}}  
 \equ 
 \nt We send the reader to these papers and move on.

\subsection{A Necessary and Sufficient Condition for USC Selections}

In order to study the existence of an equilibrium, one does not need full continuity: a nice selection of the relevant preference relation, or mapping,  is sufficient. It is this observation that motivates the results in the literature, and in line with it, recent results on the existence of an equilibrium in games and in economies require preferences to have ``nice" selections instead of having an open graph or open fibers; see  \citet{hy16} for a most recent illustration. To be sure, 
a nice selection of a correspondence is not a new idea. \citet{mi51, mi56} provides sufficient conditions for the existence of a continuous selection of a correspondence. \citet{gm75},  Y-Chapter and YP-Paper bring this result to economics\footnote{\citet{mi56} proves a selection theorem for lsc and convex-valued correspondences whose range is a separable Banach space. YP-Paper proved a version of this theorem for convex-valued correspondences which have open fibers whose range is a topological vector space. In their result the range is a  more general space but their continuity assumption is stronger.} and provide results on existence of a maximal element and an equilibrium in games and economies with possibly discontinuous preferences. In this line of literature, recent works assume weaker conditions such as constant local selection, continuous local selections and finally upper semi-continuous local selections. Our selection result provides a necessary and sufficient condition for the existence of such nice selections, hence a complement to these results.\fn{The selection has also convex-valued in the literature. We do not assume a linear structure on the space, hence convexity is not relevant. In $\Re^n$, however, convex hull of an usc correspondence is usc, hence suitable convexity assumptions on preferences, convex-valued and usc selections exist.}

\df
Let $E,X$ be two topological spaces and $M:E\tra X$ be a correspondence. 
We say $M$ is {\bf closed} if its graph, \text{\nf Gr}$M$, is closed  in $E\times X$.
\lb{df: usc}
\dff

Although the upper semi-continuity and closedness of a correspondence are distinct topological properties, they are equivalent under some suitable assumption on the range of the correspondence or the correspondence itself; see \citep[Section 17.2]{ab06}.

We next define a selection and a local selection of a correspondence. 
\df
Let $E,X$ be two topological spaces, $M:E\tra X$ a correspondence and  $E_M=\{e\in E| M(e)\neq\emptyset\}$.  
\ben[{\nf (i)},  topsep=3pt]
\setlength{\itemsep}{-1pt} 
 \ml  A {\bf selection of $M$} is a correspondence $\psi:E\tra X$ such that \text{\nf Gr}$\psi\subseteq \text{\nf Gr}M$ and $\psi(e)\neq \emptyset$ for all $e\in E_M$.  
 \ml  A {\bf local selection of $M$ at $a\in E$} is a pair $(V^a, \psi^a)$ of an open neighborhood $V^a$ of $a$ and a correspondence $\psi^a:V^a\tra X$ such that  $\text{\nf Gr}\psi^a\subseteq \text{\nf Gr}M$ and $\psi^a(e)\neq \emptyset$ for all $e\in V^a\cap E_M$. We say a local selection $(V^a, \psi^a)$ is {\bf closed} if $ \text{\nf Gr}\psi^a$ is closed in the subspace $V^a\times X$. 
\ml We say $M$ has {\bf local selections} if it has a local selection at all $a\in E_M$. 
\een
\lb{df: lselection}
\dff

 We now introduce a new continuity assumption on a correspondence. It is considerably weaker than both upper semi-continuity and closed graph properties.

\df
 Let $E,X$ be two topological spaces. A correspondence $M:E\tra X$ has the  
 {\bf neighborhood selection property (NSP)}  if 
  every $a\in E_M$ has an open neighborhood $V^a$  such that for all $e\in V^a$ and all $x\notin M(e)$ there exists an open neighborhood $U$ of $(e,x)$  in $V^a\times X$ such that  $\{e'\}\times M(e')\nsubseteq U$ for all $(e',x')\in U\cap (E_M\times X)$.  
\lb{thm: nsp}
\dff

\noindent Note that if a correspondence is closed, then in the definition above $(\{e'\}\times M(e'))\cap U^{(e,x)}=\emptyset$ for all $(e',x')\in U^{(e,x)}$. Therefore, any closed correspondence satisfies the NSP. However, the converse argument is not true -- the first example below illustrates this. The second example illustrates a correspondence which  does not satisfy the NSP.

\ex  {\nf Let $E=X=[0,1]$ be endowed with the usual metric. The correspondence $M:E\tra X$ is defined as $M(e)=(e,1)$ if $e<1$ and  $M(1)=\emptyset.$ It is clear that $M$ is not closed. The correspondence $\xi:E\tra X$ defined as $\xi(e)=\{0.5+0.5e\}$ if $e<1$ and  $\xi(1)=\emptyset$ is a selection of $M$ which is closed in $[0,1)\times [0,1].$ Moreover, the correspondence $M$ has closed local selections and has the NSP. }
\exx 

\ex {\nf Let $E=X=[0,1]$ be endowed with the usual metric. The correspondence $M:E\tra X$ is defined as $M(e)=1$ if $e$ is rational and  $M(e)=0$ otherwise. It is clear that $M$ has no closed (local) selections since $M$ has singleton values and its graph is not closed. It is easy to see that $M$ does not have the NSP.  }
\exx


Before we present our results, it is useful to observe that one can equivalently define the NSP as  follows. 

\obs
Let $E,X$ be two topological spaces and $M:E\tra X$ a correspondence. 
We call $M$ has the {\it neighborhood selection property (NSP)} if $M$ has a local selection $(V^a, \psi^a)$ at each $a\in E$ with $M(a)\neq \emptyset$ such that for all $(e,x)\in V^a\times X,$ $x\notin M(e)$ implies there exists an open neighborhood $U^{(e,x)}$ of $(e,x)$ in $V^a\times X$ such that $x'\notin \psi^a(e')$ for all $(e',x')\in U^{(e,x)}$. 
\lb{obs: nsp}
\obss

\nt Note that no topological assumption is imposed on the correspondence $\{\psi^a \}$, hence it does not necessarily be closed. In order to see this note that for those $(e',x')\notin \psi^a$ which satisfy $(e',x')\notin M$,  the existence of $U^{(e,x)}$ implies that $U^{(e,x)}\cap \psi^a=\emptyset$. However, for those $(e',x')\notin \psi^a$    which satisfy $(e',x')\in M$, we do not have open neighborhoods. Hence, $\psi^a$  need not be a closed correspondence. 

The selection properties in the literature assume the existence of a closed  (local) selection.  The following proposition provides necessary and sufficient conditions for the existence of a closed local selection of a correspondence.

\prp
Let $E,X$ be two topological spaces and $M:E\tra X$ a correspondence.  Then, the following are equivalent.
\ben[{\nf (i)},  topsep=3pt]
\setlength{\itemsep}{-1pt} 
 \ml $M$ has closed local selections. 
 \ml $M$ has the NSP.
 \een
If $M(E)$ is compact and Hausdorff, then the local selections are usc. 
\lb{th: usclselection}
\prpp

\prf[Proof of Proposition \ref{th: usclselection}]   
 Let $E,X$ be two topological spaces and $M:E\tra X$ a correspondence. 
\medskip

\noindent {\it Sufficiency.} Pick $a\in E$ such that $M(a)\neq\emptyset.$ Assume $(V^a, \xi^a)$ is a closed local selection of $M$ at $a.$  Then $\xi^a$ has a closed graph in $V^a\times X$, $\text{\nf Gr}\xi^a\subseteq \text{\nf Gr}M$ and $\xi^a(e)\neq\emptyset$  for all $e\in E_M.$  Pick $(e,x)\in V^a\times X$ with $x\notin M(e)$. Then, $(e,x)\notin \text{\nf Gr}\xi^a.$ Since $(\text{\nf Gr}\xi^a)^c$ is open in $V^a\times X$, there exists an open neighborhood $U^{(e,x)}$ of $(e,x)$ in $V^a\times X$ such that $x'\notin \xi^a(e')$ for all $(e',x')\in U^{(e,x)}$. Therefore, $M$ has the NSP. 
 
\medskip
 
\noindent {\it Necessity.} Assume $M$ has the NSP.  Pick $a\in E_M$.   Observation \ref{obs: nsp} implies that $M$ has a local selection $(V^a, \psi^a)$ at $a$ such that for all $(e,x)\in V^a\times X,$ $x\notin M(e)$ implies there exists an open neighborhood $U^{(e,x)}$ of $(e,x)$ in $V^a\times X$ such that $x'\notin \psi^a(e')$ for all $(e',x')\in U^{(e,x)}$. Define $\mathcal U^{(a,X)}=\{U^{(e,x)}| e\in V^a, x\notin M(e)\}$ and $U^{(a,X)}= \bigcup \mathcal U^{(a,X)}$. 

We next define a closed local selection $\xi^a:V^a\tra X$ of $M$ at $a$ as Gr$(\xi^a)=\left(U^{(a,X)}\right)^c$.  Then $\xi^a$ has a closed graph in $V^a\times X$, hence it is a  closed correspondence. It is clear that  $\left(U^{(a,X)}\right)^c\subseteq \text{\nf Gr}M,$ hence $\xi^a(e)\subseteq M(e)$ for all $e\in V^a.$ Assume $\xi^a(e_0)=\emptyset$ for some $e_0\in V^a\cap E_M$. By construction of $\xi^a,$ for all $x\in X,$  $(e_0,x)\in U$ for some $U\in \mathcal U^{(a,X)}.$  Then, $x\notin \psi^a(e_0)$ for all $x\in X,$ i.e., $\psi^a(e_0)=\emptyset$ while $M(e_0)\neq\emptyset.$ This furnishes us a contradiction with $\psi^a$ being a local selection of $M$ at $a$. 
Therefore, $\xi^a$ is a closed local selection of $M$ at $a.$ This completes the necessity argument. 

Finally, the upper semi-continuity of the local selections $(V^a, \xi^a)$ defined on the subspace $V^a$ follows from the Closed Graph Theorem \citep[Theorem 17.11]{ab06}.
\prff

Proposition \ref{th: usclselection} is on the local selections of a correspondence. The next result extends the local selections to global selections under the paracompactness hypothesis. This extension of ``nice" local properties to a ``nice" global property can be found in the literature for topological vector spaces; see for example \citet[Theorem 3.1$'''$]{mi56}, \citet[Theorem 3.1]{yp83}, \citet[Theorem 1]{ws96} and \citet[Lemma A. 2]{bs09}. The following proposition shows that when only the topological properties are relevant, this extension result is true for purely topological spaces. Moreover, the following theorem also relates it to the NSP which is the novel contribution of this paper. 

\prp
Let $E,X$ be two topological spaces and $M:E\tra X$ a correspondence such that $E_M$ is paracompact and Hausdorff.\fn{Every regular and Lindelof space is  paracompact and Hausdorff, hence we can replace the latter with the former assumptions (Munkres, Theorem 41.5, p. 257).  Similarly, every subspace of a metric space is paracompact and Hausdorff.} Let $M|_{E_M}$ denote the restriction of $M$ on the subspace $E_M$. Then, the following are equivalent.
\ben[{\nf (i)},  topsep=3pt]
\setlength{\itemsep}{-1pt} 
 \ml $M|_{E_M}$ has closed local selections.
 \ml $M|_{E_M}$ has the NSP.
 \ml $M|_{E_M}$ has a closed selection.
 \een
If $M(E)$ is compact and Hausdorff, then the selections are usc. 
\label{th: uscgselection}
\prpp

\prf[Proof of Proposition \ref{th: uscgselection}]   
 Let $E,X$ be two topological spaces and $M:E\tra X$ be a correspondence such that $E_M$ is paracompact and Hausdorff. 
 The proof of $M|_{E_M}$ has closed local selections if and only if it has the NSP follows from Proposition \ref{th: usclselection}. It remains to prove that $M|_{E_M}$ has a closed selection if and only if it has closed local selections.

\smallskip

\noindent {\it Sufficiency.} Assume $\xi: E_M\tra X$ is a closed selection of $M|_{E_M}$.  Then, $\xi$ has a closed graph in $E_M\times X$, non-empty values and $\xi(e)\subseteq M(e)$ for all $e\in E_M.$  Therefore $(E_M,\xi)$ is a closed local selection of $M|_{E_M}$ for all $a\in E_M,$. 
 
\medskip
 
\noindent {\it Necessity.} For all $a\in E_M,$ assume $(V^a, \xi^a)$ is a closed local selection of $M|_{E_M}$ at $a.$ We now construct a closed selection $\xi$ of $M|_{E_M}$. Note that the collection $\mathcal V^a=\{V^a\}_{a\in E_M}$ is an open cover of $E_M.$ Since $E_M$ is paracompact and Hausdorff,  $\mathcal V$ has a closed, locally finite refinement $\mathcal K=\{K^{a_\alpha}\}_{\alpha\in D}$ which covers $E_M$ \citep[Lemma 1]{mi53}, i.e., $\mathcal K$  is a covering of $E_M$ consisting of closed sets such that every element of $\mathcal K$  is a subset of  some element of $\mathcal V,$ and each $a\in E_M$ has an open neighborhood which intersects only finitely many elements of $\mathcal K$.  It follows from $K^\alpha\times X$ is closed in $E_M\times X$ for all $\alpha\in D$ and the union of a locally finite collection of closed set is closed \citep[Lemma 39.1]{mu00} that the correspondence $\xi:E_M\tra X$ defined as
$$
\mbox{Gr$\xi$}=\bigcup_{\alpha\in D}\mbox{Gr$\xi^{a_\alpha}$}\cap \left(K^{a_\alpha}\times Y\right)
$$
has a closed graph in the subspace $E_M\times X$. Since $\xi(a)\subseteq M(a)$ for all $a\in E_M,$ $M|_{E_M}$ has a closed selection.

Finally, the upper semi-continuity of the selections defined on the subspace $E_M$ follows from the Closed Graph Theorem \citep[Theorem 17.11]{ab06}.
\prff

\section{Applications to Walrasian Equilibrium Theory}

In this section we present four applications of the results exposed in the above section to underscore the wide-range applications of of Yannelis' selection theorem.  three propositions on Berge’s theorem, a generalization o the GND lemma and  another on the Browder-McCabe-Yannelis map, an existence  result in Shafer’s setting of a non-transitive consumer, and finally, a generalization  the approximation result of Anderson-Khan-Rashid to preferences that are not necessarily continuous. We emphasize that like \cite{co20}, we remain in a finite-dimensional setting.

\subsection{On Berge's Maximum Theorem}

There are two major applications of Berge's maximum theorem: (i) existence of an equilibrium and (ii) the structure of the equilibrium correspondence. For the latter, the upper semi-continuity of the entire argmax correspondence is a desirable property. However, as mentioned in Section 2.4, for the existence of an equilibrium, a nice selection is sufficient. Berge's theorem shows that if a function $u$ is continuous on $E\times X$ and the constraint correspondence $F: E\tra X$ is both upper and lsc and has non-empty, compact values, then the $\argmax$ correspondence has non-empty and compact values,  and is usc.  The literature on the generalization of the Berge's maximum theorem is preoccupied with the upper semi-continuity of the $\argmax$ correspondence. \citet{wa79} provides a general version of Berge's theorem for preferences (and for abstract correspondences). \citet{dm89} weaken continuity of the objective function but impose order theoretic and convexity assumptions on it. \citet{tz95} provide a necessary and sufficient condition for the upper semi-continuity and the non-empty-valuedness of the argmax correspondence.\fn{See also \cite{ms07b} for a recent generalization of Berge's theorem. Moreover, in this paper we explicitly focus on topological generalizations of Berge's theorem. There is a literature on measure-theoretic version of Berge's theorem  which we do not pursue; see   \cite{cv77} and \cite{kh85} for details.}  In this subsection we present three propositions weakening the continuity assumptions of the Berge's maximum theorem, one for functions and the other two for correspondences, which provide sufficient conditions on the 
existence of an usc selection for the argmax correspondence.

We start with functions.  Let $E,X$ be two topological spaces, $F:E\tra X$ a correspondence and $u:E\times X\ra \Re$ a function.  Define $M:E\tra X$ as 
$$
M^u(e)=\argmax_{x\in F(e)} u(e,x). 
$$


\df
Let $E, X$ be two topological spaces, $F: E\tra X$ a correspondence and $u: E\times X\ra \Re$ a function. We say $u$ has the {\bf continuous selection property with respect to $F$} if every $a\in E$ has an open neighborhood $V^a$ such that for all $(e,x)\in \text{\nf Gr}F\cap (V^a\times X)$ with $x\notin M^u(e)$, there exist an open neighborhood $U^{(e,x)}$ of $(e,x)$ in $V^a\times X$ and $y\in X$  such that for all $(e',x') \in U^{(e,x)}$, $u(e', y)\geq u(e', x')$  and $(e',y)\in\text{\nf Gr}F\backslash U^{(e,x)}$.
\label{dfn: nsputility}
\dff

\nt It is easy to observe that the continuous selection property is considerably weaker than those used in \citet[Definitions 7, 8, 9]{tz95}, severally, and collectively, hence the following generalizes \citet[Theorem 3]{tz95}. 

\prp
Let $E,X$ be two topological spaces, $E$ paracompact and Hausdorff, $F:E\tra X$ a correspondence with non-empty and compact values and $u:E\times X\ra \Re$ a function that has the continuous selection property with respect to $F$.  Then $M^u$ has non-empty values and a closed selection.  If, in addition, $F(E)$ is compact and Hausdorff, then the selection is usc. 
\lb{thm: bergeu}
\prpp

\prf[Proof of Proposition \ref{thm: bergeu}]  
First, it is easy to observe that the non-emptiness of $M^u(e)$ for all $e\in E$ follows from \citet[Theorem 1]{tz95}.   
Second, since $u$ has the  continuous selection property  with respect to $F$,  every $a\in E$ has an open neighborhood $V^a$ such that for all $(e,x)\in \text{\nf Gr}F\cap (V^a\times X)$ with $x\notin M^u(e)$, there exist an open neighborhood $U^{(e,x)}$ of $(e,x)$ in $V^a\times X$ and $y\in X$  such that for all $(e',x') \in U^{(e,x)}$, $u(e', y)\geq u(e', x')$  and $(e',y)\in\text{\nf Gr}F\backslash U^{(e,x)}$. Now pick $(e',x') \in U^{(e,x)}$. If $y\in M^u(e')$, then $\{e'\}\times M^u(e')\nsubseteq U^{(e,x)}$. If $y\notin M^u(e')$, then pick $\hat y\in M^u(e')$. It follows from $u(e', \hat y)>u(e', y)\geq u(e', x')$ that $\{e'\}\times M^u(e')\nsubseteq U^{(e,x)}$. Therefore, $M^u$ has the NSP. Then Proposition \ref{th: uscgselection} imply that $M^u$ has a closed selection, and when $F(E)$ is compact and Hausdorff, then this selection is usc.  
\prff

Note that Proposition \ref{thm: bergeu} suggests that the continuous selection property  provided in Definition \ref{dfn: nsputility} allows the $\argmax$ correspondence to have a graph which is not closed. Hence,  it does not rule out the possibility that a point $x$ is not a maximal element at some parameter $e$, yet all of its neighborhoods contains maximal elements. 

We next turn to the generalization of Berge's theorem for correspondences. Let $E,X$ be two topological spaces, $F:E\tra X$ and $P
:E\times X\tra X$ be two correspondences.    
Define a correspondence $M^P: E\tra X$ as 
$$
M^P(e)=\{x\in F(e)~|~P(e,x)\cap F(e) =\emptyset\}.
$$

\nt We say the correspondence $P$ is  {\it  irreflexive} if $x\notin P(e,x)$ for all $(e,x)\in E\times X$, and {\it fully transitive} if for all $e\in E$ and all $x,y,z\in X,$  $y\notin P(e, x)$ and $z\notin P(e, y)$ implies $z\notin P(e, x)$.\fn{See \cite{zt92} and \cite{ku19a} for a detailed discussion on different transitivity postulates. Note that the irreflexivity and full transitivity of a correspondence are not unusual properties if we interpret the graph of $P(e,\cdot)$ as a binary relation on $X$.}  


\smallskip

\df
Let $E, X$ be two topological spaces, $F: E\tra X$ a correspondence and $P: E\times X\ra X$ a function. We say $P$ has the {\bf continuous selection property with respect to $F$} if every $a\in E$ has an open neighborhood $V^a$ such that for all $(e,x)\in \text{\nf Gr}F\cap (V^a\times X)$ with $x\notin M^P(e)$, there exist an open neighborhood $U^{(e,x)}$ of $(e,x)$ in $V^a\times X$ and $y\in X$  such that for all $(e',x') \in U^{(e,x)}$, $x'\notin P(e', y)$  and $(e',y)\in\text{\nf Gr}F\backslash U^{(e,x)}$.
\label{dfn: nsppreference}
\dff

\nt Analogous to the discussion above, the continuous selection property (for correspondences) is weaker than those used in \citet[Definitions 10, 11, 12]{tz95}, severally, and collectively, hence the following result weakens the continuity assumption of  \citet[Theorem 4]{tz95}. 

\prp
Let $E,X$ be two topological spaces, $E$ paracompact and Hausdorff, $F:E\tra X$ a correspondence with non-empty and compact values and $P:E\times X\tra X$ an irreflexive and fully transitive correspondence that has the continuous selection property with respect to $F$.  Then $M^P$ has non-empty values and a closed selection.  If, in addition, $F(E)$ is compact and Hausdorff, then the selection is usc. 
\lb{thm: bergep}
\prpp

\prf[Proof of Proposition \ref{thm: bergep}]  
First, it is easy to observe that the non-emptiness of $M^P(e)$ for all $e\in E$ follows from \citet[Theorem 1]{zt92}.   
Note that it follows from \citet[Proposition 2]{ku19a} that $P$ is transitive, i.e.,  for all $e\in E$ and all $x,y,z\in X,$  $z\in P(e, y)$ and $y\in P(e, x)$ implies $z\in P(e, x)$. Second, since $P$ has the continuous selection property with respect to $F$,  every $a\in E$ has an open neighborhood $V^a$ such that for all $(e,x)\in \text{\nf Gr}F\cap (V^a\times X)$ with $x\notin M^P(e)$, there exist an open neighborhood $U^{(e,x)}$ of $(e,x)$ in $V^a\times X$ and $y\in X$  such that for all $(e',x') \in U^{(e,x)}$, $x'\notin P(e', y)$  and $(e',y)\in\text{\nf Gr}F\backslash U^{(e,x)}$. 

Now pick $(e',x') \in U^{(e,x)}$. If $y\in M^P(e')$, then $\{e'\}\times M^P(e')\nsubseteq U^{(e,x)}$. If $y\notin M^P(e')$, then  there exists $z\in F(e')$ such that $z\in P(e',y)$. If $z\in M^P(e')$, then transitivity and irreflexivity of $P$ imply $\{e'\}\times M^P(e')\nsubseteq U^{(e,x)}$. Otherwise, there exists $z'\in F(e')$ such that $z'\in P(e',z)$. Now pick $\hat y\in M^P(e')$. Then, it follows from full transitivity of $P$ that $\hat y \in P(e',z)$ (otherwise $z'\notin P(e',\hat y)$ and $\hat y \notin P(e',z)$ yield a contradiction). Then transitivity  of $P$, $\hat y \in P(e',z)$ and $z \in P(e',y)$ imply $\hat y \in P(e',y)$.  Hence  transitivity and irreflexivity of $P$ imply $\{e'\}\times M^P(e')\nsubseteq U^{(e,x)}$.  Therefore, $M^u$ has the NSP. Then Proposition \ref{th: uscgselection} implies that $M^u$ has a closed selection, and when $F(E)$ is compact and Hausdorff, then this selection is usc.  
\prff

It is well-known in mathematical economics that without the transitivity assumption, topological assumptions on a choice set and on a preference relation defined on it are not enough to guarantee the existence of a maximal element. However, with the added linear structure on the choice set and a suitable convexity assumption on preferences, the existence of a maximal element is guaranteed. The following result provides a maximum theorem with a weak continuity assumption on preferences which is not necessarily complete or transitive but satisfies a convexity assumption.  

  \prp
Let $E$ be a non-empty topological space, $X$ a non-empty  Hausdorff locally convex topological vector space,  $F:E\tra X$ a closed correspondence with non-empty, convex and compact values,  $P:E\times X\tra X$ a correspondence with $x \notin \text{\nf co}P(e, x)$ for all $(e,x)\in E\times X$, and  the correspondence $\psi: E\times X\tra X$ defined as $\psi(e,x)=P(e,x)\cap F(e)$ has the continuous inclusion property.  Then $M^P$ is a closed correspondence with non-empty values.  If, in addition, $F(E)$ is compact, then $M^P$ is usc. 
\lb{thm: bergepc}
\prpp

\prf[Proof of Proposition \ref{thm: bergepc}]
It follows from \citet[Corollary 1]{hy17} that for each $e\in E$, $M^P(e)\neq\emptyset$.  It follows from $\psi$ having the continuous inclusion property that $P$ has quasi-transfer open lower sections in $(e,x)$ with respect to $F$ \citep[Definition 10]{tz95}. Then the argument in the proof of  \citet[Theorem 4]{tz95} shows that $M^P$ is closed. Finally, when $F(E)$ is compact,  the upper semi-continuity of $M^P$ follows from the Closed Graph Theorem \citep[Theorem 17.11]{ab06}.
\prff

\subsection{The Gale-Nikaido-Debreu Lemma}

Inspired by the development in discontinuous games and economies prior to the work of \cite{co20}, in this subsection we present a  generalization of the GND lemma by imposing a weak continuity assumption on Browder-McCabe-Yannelis map -- the continuous inclusion property.   \citet[Theorem 1]{mc81a} and \citet[Theorem 3.1]{ya85} assumed the excess demand correspondence is usc, \citet[Theorem 4]{hy17} and \citet[Theorems 1 and 2]{co20} assume it has the continuous inclusion property, hence the Browder-McCabe-Yannelis map  has open fibers. 
\citet[Theorem 8]{mt87},   on the other hand,  impose a direct assumption on the Browder-McCabe-Yannelis map, and assumed that it has the local intersection property.  Proposition \ref{thm: GND} below assumes a weaker continuity assumption on the Browder-McCabe-Yannelis map, hence generalizes these results as well as previous results cited in these papers for finite dimensional commodity spaces.\fn{Note that Yannelis' work on the GND lemma is on infinite dimensional spaces. We limit ourselves to finite dimensional commodity spaces. Proposition  \ref{thm: GND} was initially presented in \citet[p. 115--116]{uy16a}.} Moreover, we show that a continuity assumption on the excess demand correspondence which put restrictions only on the {\it downward jumps,} hence weaker than upper semi-continuity, which implies the Browder-McCabe-Yannelis map has the continuous inclusion property.  



An economy $\mathcal E= (Z, \zeta)$ with $m$ commodities is defined as follows:  $Z\subseteq \Re^m$ is the set of possible excess demands which is non-empty, convex and compact, and $\zeta:\Delta\tra Z$ the excess demand correspondence where $\Delta= \{p\in \Re_+^m |~\sum_{k=1}^m p^k=1\}$ is the set of prices.  Let $-\Omega= \{x\in \Re^m |~x\leqq 0\}.$ The {\it Browder-McCabe-Yannelis map} is a correspondence  $\Psi:\Delta\tra \Delta$ defined as 
$$\Psi(p)=\{q\in \Delta: q\cdot \zeta(p)>0\}.$$


\prp
Let  $\mathcal E= (Z, \zeta)$ be the economy defined above such that 
\ben[{\nf (i)}, topsep=0in]
  \setlength{\itemsep}{0in}
  \setlength{\parskip}{0pt}
  \setlength{\parsep}{0pt}

\ml the Browder-McCabe-Yannelis map $\Psi$ has the continuous inclusion property, 
\ml $\zeta$ has non-empty, convex and closed values,
\ml for each $p\in \Delta$ there exists $x\in \zeta(p)$ such that $p\cdot x \leqq 0.$ 
\een
Then there exists $p^*\in \Delta$ such that $\zeta(p^*)\cap -\Omega\neq \emptyset.$
\lb{thm: GND}
\prpp

\prf[Proof of Proposition \ref{thm: GND}]  
Assume the conclusion of Proposition \ref{thm: GND} is false. Then $\zeta(p)\cap -\Omega= \emptyset$ for each $p\in \Delta.$ Also, since $\zeta(p)$ is non-empty, compact, convex set and $-\Omega$ is a closed, convex set, there exists $q\in \Delta$ that strictly separates $\zeta(p)$ and $-\Omega$ for each $p\in \Delta.$ Hence, the correspondence $\Psi$ has non-empty values. It is clear that $\Psi$ has convex values. And since $\Psi$ is assumed to have the continuous inclusion property,  Theorem (He-Yannelis) implies there exists $p^*\in \Delta$ such that $p^*\in \Psi(p^*),$ i.e. $p^*\cdot x>0$ for all $x\in \zeta(p^*).$ This furnishes a contradiction with (iii). 
\prff

In Proposition \ref{thm: GND}, we assume the Browder-McCabe-Yannelis map $\Psi$ has the continuous inclusion property which is merely a technical assumption. Now, we propose a weak continuity assumption on the excess demand correspondence which imposes a restriction only on downward jumps and implies that $\Psi$  has the continuous inclusion property.

\df 
Let $X, Y$ be  two non-empty subsets of $\Re^m$ and $P:X\tra Y$ a correspondence. $P$ is {\bf continuous from below (cfb) at $x\in X$} if for each open half-space $H$ containing he closure of $P(x)$, 
there exists an open neighborhood $U$ of $x$ such that for all $x'\in U$ and all $z'\in P(x')$, there exists $z\in H$ such that $z\leqq z'.$  The correspondence $P$ is {\bf cfb} if it is cfb at each $x\in X.$ 
\label{dfn: cfb}
\dff


\noindent 
It is easy to see that cfb is weaker than upper semi-continuity. Now, let $X, Y$ be non-empty subsets of $\Re^m,$ $P:X\tra Y$ a correspondence, and $\pi$ the usual projection map. Define $\pi^k P: X\tra \Re$ as $\pi^k P(x)= \pi^k (P(x))$ and $\underline{x}^k= \mbox{ min } \pi^k P(x)$ for each $k=1, \ldots, m.$ 


\df  
Let $X, Y$ be non-empty subsets of $\Re^m$ and $P:X\tra Y$ a correspondence. $P$ is {\bf weakly continuous from below (wcfb) at $x\in X$} if {\nf (i)} $K=\{k: \underline{x}^k>0, k=1,\ldots,m\}\neq\emptyset$ implies $\pi^k P$ is cfb at $x$ for some $k\in K$, and {\nf (ii)} $K=\emptyset$ implies $P$ is cfb at $x.$ The correspondence  $P$ is  {\bf wcfb} if it is wcfb at each $x\in X.$ 
\label{dfn: wcfb}
\dff


\noindent Note that, if at some price $p,$ the excess demand of at least one commodity is positive and its excess demand remains positive in a neighborhood of $p,$ then this is sufficient for the excess demand correspondence to be wcfb at $p,$ irrespective of the behavior of the excess demand of other commodities around $p.$ 


\prp
Let  $E= (Z, \zeta)$ be the economy defined above such that $\zeta$ is wcfb and has non-empty, closed values. Then the Browder-McCabe-Yannelis map $\Psi$ has the continuous inclusion property.
\lb{thm: GNDcns}
\prpp


\prf[Proof of Proposition \ref{thm: GNDcns}] 
First, it follows from $Z$ is compact and $\zeta$ has non-empty and closed values that $\pi^k \zeta$ has non-empty and compact values, and hence, $\underline{p}^k= \mbox{ min } \pi^k \zeta(p)$ is well defined for each $k=1, \ldots, m$.  
Now pick $p\in \Delta$ such that $\Psi(p)\neq\emptyset$. Then,  there exists $q\in \Delta$ such that $q\cdot \zeta(p)>0.$ 

First, assume for some $k=1,\ldots, m,$ $\underline{p}^k>0$ and $\pi^k \zeta$ is cfb at $p.$ Since $\zeta(p)$ and $Z$ are compact, $\pi^k \zeta(p)\times \pi^{-k} (Z)$ is compact. Then, there exists $\bar q\in \Delta$ that strictly separates $\pi^k \zeta(p)\times \pi^{-k} (Z)$ and $-\Omega.$  And, since $\pi^k\zeta$ is cfb at $p\in \Delta,$ for sufficiently small  $\varepsilon>0,$ there exists an open neighborhood $U^p$ of $p$ such that for all $p'\in U^p,$ $\bar  q$ still strictly separates $\{\pi^k \zeta(p)-\varepsilon\}\times \pi^{-k} (Z).$ Then $\bar q \cdot \zeta(p')>0$ for all $p'\in U^p$. Therefore, $\Psi$ has the local intersection property at $p$. 
Second, assume for each commodity $k=1,\ldots, m,$ $\underline{p}^k\leqq 0.$ Then $\zeta$ is cfb at $p$. And since $q\cdot \zeta(p)>0,$ $q$ determines an open half space $H^q$ in $\Re^m$ containing $\zeta(p).$ And since $\zeta$ is $cfb,$ there exists an open neighborhood $U^p$ of $p$ such that for all $p'\in U^p$ and all $z'\in \zeta(p')$, there exists $z\in H^q$ such that $z\leqq z'.$ Therefore, $q\cdot \zeta(p')>0$ for all $p'\in U^p,$ hence $\Psi$ has the local intersection property at $p.$ 

Therefore, it follows from $\Psi$ has the local intersection property at all $p\in \Delta$ that $\Psi$ has the continuous inclusion property.  
\prff


We end this section by providing some examples of aggregation of correspondences.  In economies, existence of an equilibrium does not require well-behaved individual demand correspondences -- well-behaved aggregate demand correspondence is enough as Uzawa suggested to \citet[note 1, p88]{de59} for production sets: individual demands as functions of prices are allowed to jump. As long as the aggregate demand has closed local selections, then we can have an (approximate) equilibrium; see Subsection 3.4 below.  

\df
Let $E,X$ be topological space. We call a correspondence $F:E\tra X$ {\bf jumps at $e\in E$} if $F$ has no closed local selection at $e.$
\lb{df: jump}
\dff

\exm {\nf Let $E=X=\Re_+$ and $F^i(e)=\{0,k\}, k>0,$ for $e\neq 1$ and $F^i(1)=\{k/n\}$ for all $i=1,\ldots,n$, $n>1$. Since $F^i$  jumps at $1$ for all $i$, it follows from Proposition \ref{th: uscgselection} that each $F^i$ does not have an usc selection. Although the aggregate correspondence $F=\sum_{i=1}^nF^i$ is not usc, it has an  usc~selection $\zeta$ defined as $\xi(e)=\{k\}$ for all $e\in E$.}
\label{exm: jump1}
\exmm


\exm {\nf  Let $E=\{p\in \Re_{+}^2|0<p\leqq 1\}$ denote the set of all strictly positive prices. Consider the following two-agent, two-good  pure-exchange economy. Let $X=\Re_+^2$ be the consumption set of both agents. The endowments are $\omega^A=\omega^B=(1,1).$ The preference of agents are such that their demand correspondences are as follows: 
\begin{align*}
x^A(p)= \left\{\begin{array}{ll}
\left\{(\frac{p_1+p_2}{p_1},0)\right\}, & \text{ if } p_1\leq p_2\\
\left\{(0,\frac{p_1+p_2}{p_2})\right\}, & \text{ if } p_1> p_2\\
\end{array}\right.
& ~\text{ and }~
x^B(p)=\left\{\left(\frac{p_1+p_2}{p_1}, 0\right), \left(0,\frac{p_1+p_2}{p_2}\right)\right\}.
\end{align*}
Agent $A$ strictly prefers good 1 over good 2 if its price is lower, however, he shifts to good 2 when the first commodity is more expensive. Agent $B$ does not like to mix.  It is clear that agent $A$'s demand correspondence jumps at prices $p$ such that $p_1=p_2$, hence it has no usc selection.  However, the aggregate demand correspondence has an usc selection $\zeta(p)=\left\{\left(\frac{p_1+p_2}{p_1}, \frac{p_1+p_2}{p_2}\right)\right\}$. 
} 
\label{exm: jump2}
\exmm


%


%
%

\subsection{On Shafer's Non-Transitive Consumer}

In the existence of a Walrasian equilibrium, two main approaches have been used in economic theory: simultaneous optimization and excess demand. The former was introduced by \cite{de52} and used in \citet{ad54} while the latter was introduced in Gale-Debreu-Nikaido and used in \citet{de59}. The former was picked up by \citet{ss75} and YP-Paper; see \cite{de82} for a detailed comparison.\fn{Of course this excludes McKenzie's and Negishi-Moore's methods; see \cite{mc81b, mc02} for a discussion.}  Without the transitivity assumption, the excess demand approach may not work since it is not guaranteed that the excess demand correspondence is convex valued; see \cite{so77}. However, if one guarantees that a nice selection exists, than this method can still work. In a resect work, \cite{sc15} uses extension of preferences in order to obtain a nice selection of the demand correspondence. \citet{gm75} use a similar extension result about half a century ago in order to obtain existence of an equilibrium in an economy.\footnote{It is of interest to explore if these two augmentations/extensions are  relevant. \citet[Definition 15]{be92} implies that \cites{gm75}   augmented preference is equivalent to Scapparone's!}

 In this subsection we provide a simple economy  where the demand correspondence is a function, and hence convex valuedness is not an issue.   
The commodity space is $X$ be non-empty, convex and compact subset of $\Re_+^k$ and $I$ denotes a finite set of individuals. For each agent $i$, let $\succcurlyeq_i\subseteq X\times X$ denote her  preference relation and $e_i\in \Re_{++}^m$ her the endowment. The set of prices is $\Delta=\{p\in \Re_+^m |~ 0\leqq p\leqq 1\}$. Define the budget correspondence of agent $i$ as $B_i(p)=\{x\in \Re_+^m: ~px\leqq pe_i\}$. Assume the preference relation of each agent $i$ satisfies the following assumptions: 

\ben[(i),  topsep=3pt]
\setlength{\itemsep}{-1pt} 
\ml the correspondence $\psi_i: \Delta\times X$ defined as  $\psi_i(p,x)=B_i(p)\cap P_i(x)$ has the continuous inclusion property, where $P_i(x)=\{y\in X:y\succ_i x\}$,  
\ml strong convexity, i.e., $x\succcurlyeq_i y$ and $z\succcurlyeq_i y$, $x\neq z$ imply $\lambda x+ (1-\lambda) z\succ_i z$ for all $x,y,z\in X$ and all $\lambda\in (0,1)$. 
\een

\nt  For each $i\in I$ and $p\in \Delta,$ the excess demand correspondence of individual $i$ is 
$$
d_i(p)=\{x-e_i|~ x\in \Re_+^m, ~px\leqq pe_i, ~y\succ_i x \Rightarrow py>pe_i\}=\{x-e_i|~ x\in B_i(p), \psi_i(p,x)=\emptyset\}.
$$

\prp
Given the exchange economy just described, the excess demand correspondence is a continuous function. 
\label{thm: shafer}
\prpp

\prf[Proof of Proposition \ref{thm: shafer}]
Proposition \ref{thm: bergepc} implies that $d_i$ is usc with non-empty values. It follows  (ii) that $M^u$ is singleton valued, hence it is continuous as a function. 
\prff

McKenzie used excess demand correspondence in the presence of non-ordered, non-convex preferences. As mentioned above,  the excess demand correspondence may not have a convex selection, see for example \cite{so77}, but who ask us to use the excess demand derived by using the original correspondence.  McKenzie worked with an extension of the original preference relation and then by some careful and delicate arguments obtained the equilibrium; see also \cite{kh93}.  \citet[p. 824]{mc81b} had already stressed the importance of excess-demand approach: 
\bqu
I think there are advantages to the use of the demand function, or correspondence, in proofs of existence, both for mathematical power and for understanding the proof. I will show how the demand correspondence may be used in a mapping of the Cartesian product of the price simplex and the social consumption set into itself whose fixed points are competitive equilibria even in the absence of the survival assumption.\fn{McKenzie continues, \lq\lq  At the equilibrium the budget sets will not be empty, the demand correspondences will be well defined and upper semi- continuous, but these conditions need not be satisfied for non-equilibrium prices. We will avoid the difficulties posed by this possibility by using an extension of the demand correspondence which reduces to the original correspondence whenever the original correspondence is well defined and non-empty. The extended demand correspondence will be well defined and non-empty for all price vectors." in this connection also see Footnotes~\ref{fn:mck1} and \ref{fn:mck2} below and the text they footnote. \label{fn:mck0} }  
\equ

\noindent This will hold us in good stead for the result of the next section.

\subsection{On Approximate Equilibrium: Starr-Arrow-Hahn}

 In this subsection, we provide a result on the  existence of an approximate equilibrium. It weakens the continuity assumption in \cite*{akr82}.  
Consider the following exchange economy. The commodity space is $\Re_+^m$ and  $I=\{1,\ldots, n\}$ denotes a finite set of individuals $n\geq m$. The endowment of agent $i$ is $e_i\in \Re_+^m$, the set of prices is $\Delta=\{p\in \Re_+^m |~ 0<p\leqq 1\}$ and the budget correspondence of agent $i$ is $B_i(p)=\{x\in \Re_+^m: ~px\leqq pe_i\}$. Let $\CP$ denote the set of preferences $\succ_i\subseteq \Re_+^{2m}$, i.e. binary relations
on $\Re_+^m$ satisfying 
\ben[(i),  topsep=3pt]
\setlength{\itemsep}{-1pt} 
\ml {\it continuity:} the correspondence $P_i: \Delta\times \Re_+^m\tra \Re_+^m$ defined as $P_i(p, x)=\{y\in \Re_+^m: y\succ_i x\}$ has  the continuous selection property with respect to $B_i$ for all $i\in I$, 
\ml {\it full transitivity:} $x \nsucc y, ~y \nsucc z \Rightarrow x \nsucc z,$
\ml {\it irreflexivity:} $x \nsucc x,$
\een
as defined in Subsection 3.1.\fn{In the result presented in this section, it is possible to weaken full transitivity by replacing it with the transitivity, or acyclicity, of $\succ$. This can be achieved by strengthening the continuity assumptions on the preference and budget correspondences; see for example \citet[Proposition 2 and Theorem 4]{tz95} for sufficient conditions to obtain a non-empty valued, upper semi-continuous argmax correspondence.}  

An exchange economy is a map 
$\mathcal E: I \ra \CP \times \Re_+^m.
$ 
For each $i\in I,$  $\mathcal E(i)=(\succ_i, e_i)$ assigns agent $i$ a preference relation $\succ_i$ and endowment $e_i$. Given $i\in I$ and $p\in \Delta,$ the excess demand correspondence of individual $i$ is 
$$
d_i(p)=\{x-e_i|~ x\in \Re_+^m, ~px\leqq pe_i, ~y\succ_i x \Rightarrow py>pe_i\}.
$$
And the excess demand correspondence $D:P\tra \Re_+^m$ is defined as $D(p)=\sum_{i\in I}d_i(p).$ Let $\| x \|$ denote $\sum_{l\ell=1}^m | x^\ell |.$ Also, for $1\leqq n'\leqq n$, let $E_{n'}=(1/n')\mbox{ max }_{S\subseteq A, |S|=n'}\| \sum_{i\in S} e_i  \|$, the norm of the average endowment of the $n'$ norm-best endowed agents, and let $E = E_n$, the norm of the economy-wide average endowment. 


\prp
Given the exchange economy just described, there exist $p\in \Delta$ and $z_i\in d_i(p)$ for all agent $i$ such that 
\vspace{-5pt}
$$ \frac{1}{n} \sum_{\ell=1}^m \mbox{ {\nf max }}\left\{\sum_{i\in I} z^\ell_i, 0\right\} \leqq 2\sqrt{\frac{m}{n} {E_mE}}.$$
\vspace{-20pt}
\lb{thm: app0}
\prpp

\prf[Proof of Proposition \ref{thm: app0}]
 It follows from Proposition \ref{thm: bergep} that the excess demand correspondence has non-empty values and a closed selection $\zeta$.  The rest of the proof  is identical to the proof of \cites{ge86} theorem (which uses a construction slightly different from the proof of the theorem in \cite*{akr82}) except that the the demand correspondence is replaced by the selection $\zeta$. 
 \prff



%
%





\section{Open Questions and Alternative Directions}
We now come to the end of this essay, and towards the completion of our portrait of Nicholas Yannelis: by necessity it is an incomplete completion. For one thing, we exclude the economics of information from our purview and do not even cite his 1991 paper that has proved so influential.  To the extent that a portrait is a metonymy for a proper name,  \cites{kr72}  \lq\lq necessary and sufficient conditions that will work for a term like reference" are intimately tied with naming and necessity, and with what  a proper name represents. Kripke writes, \lq\lq I want to present a {\it better picture}  than the picture presented by the received views."\fn{For the statement, see page 92 and  for the  context of the statement, see pages 91-97, and more generally the text to the entry \lq\lq Proper names, correct picture of reference." To doubleback to Footnote~\ref{fn:ke},  Kemp advocates the \lq\lq historian's version of
Occam's razor" regarding representation and perspective: \lq\lq In attempting to resolve this impasse the historian has, I believe, no choice but to adopt
the simplest, clearest, most direct solution which is compatible with the evidence, asking
rigorously at each stage if his solution is necessary and essential, no matter how intellectually
attractive it may be (p. 143)."   } It is in this way that we would like the work above to be read. Indeed, when one thinks of the name Nicholas Yannelis, one thinks of Lionel McKenzie and Gerard Debreu, on the one (substantive) hand, and of  infinite-dimensions, of continuous selections, of non-convex correspondences on the (technical) other.  In this concluding section, we briefly consider these signature terms. 

We begin with the technical register: with infinite-dimensionality. 
The reader has already noted perhaps a dissonance in the essay in that Section 2 on the mathematical contribution is constantly striving for a view of the subject that is not obscured  by the finite-dimensional limitation, be it of the cardinality of the space of agents or the space of commodities or actions at their disposal. This \lq\lq striving" is a subtext of the 1983 dissertation. And yet, our Section 3 on the  applications to Walrasian equilibrium theory are all set in the context of a
 finite-dimensional commodity space, even  though we have referred in passing to \cite{ya85} and \cite{hy16, hy17} in the context of the GND lemma.\fn{Footnotes~\ref{fn:fin0} and  \ref{fn:fin1}  above, and the text that they footnote,  concern these references. \label{fn:fin2}}   But surely more remains to be done, and it is this is one  direction that is clearly opened by our consideration. But there is also  another  affiliated  investigation that derives from the fact that in infinite dimensional spaces, the  convex hull of a closed set and that of a compact set  is not necessarily closed. This slips from infinite-dimensionality to non-convexity, again a preoccupation of the dissertation and the work that followed it. And indeed in the context of selection,  when the correspondence has non-convex values, one may explore the direction of ensuring nice selections from correspondences with \lq\lq star-shaped" values,  as in  
 \cite{ma79} and  \cite{mi99}.

Turning from the ideas to the names, the  signature terms for Lional McKenzie, if projected to the register of \lq\lq classical   general equilibrium theory" are two: {\it irreducibility} and the {\it survival} assumption. As regards the first, it is best to let him speak in his own words \cite[p. 374]{mc99}: 
\bqu  
One of my chief contributions to general equilibrium, following a suggestion from David Gale, is the concept of irreducibility for a competitive economy.  Loosely speaking, irreducibility means that the economy cannot be divided into two groups where one group has nothing to offer the other group which has value for it.This replaces the assumption used by Arrow and Debreu that everyone owns a factor that is always able to increase the output of a good which is always desired in larger quantities by everyone. This assumption implies irreducibility, but it is rather implausible.\fn{McKenzie continues, \lq\lq This idea was first defined and used in my paper on the existence of equilibrium published in 1959, where various generalizations were made of the theory announced by Arrow and Debreu and myself in 1954 and by me in 1956." We send the interested reader to this text and to \cite{kh20}, and also to Footnote~\ref{fn:mck0}.  \label{fn:mck1} }  \equ

\noindent As to the \lq\lq survival assumption" and to the words of  \cite[p. 823]{mc81b}: 
\bqu
Perhaps the most dramatic innovation since 1959 is the discovery that the survival assumption [...] can be dispensed with in the presence of the other assumptions, in particular in the presence of Assumption (6) that the economy is irreducible. 
\equ

\noindent And as he was to recapitulate in  \cite[p. 374]{mc99}: 
\bqu 
 It was first seen from the work of John Moore (1975) that irreducibility made this assumption unnecessary, although he did not call attention to the generalization. He applied a fixed point theorem to a mapping of a set of normalized utility possibility vectors into itself in the manner of Negishi (1960) and Arrow and Hahn (1971). This suggested that the survival assumption for isolated consumers had been needed only because the mappings were defined by demand functions in the commodity and price spaces. I was able to confirm this in my presidential address to the Econometric Society (1981).\fn{He did so  by \lq\lq showing that demand functions based on the pseudo-utility functions of Shafer with some rather difficult indirect arguments allowed a commodity space approach to the existence proof where survival for isolated individuals is not assumed. The new approach was essential in order to achieve the further generalization of the existence theorem without individual survival to the case of intransitive preferences, since in the absence of transitivity the utility functions do not exist. Thus, a mapping in the space of utility vectors is not available."  \cite[p. 208]{mc02} added \lq\lq Moreover it has been shown in the activities model (McKenzie 1981) that the survival assumption can be dropped when irreducibility is assumed. However, the argument is too involved to be introduced here."  Also see Footnotes~\ref{fn:mck0} and \ref{fn:mck1}. \label{fn:mck2}} \equ

 In the introduction to this portrait, we had occasion to refer to the impact of McKenzie and Metakides on Yannelis' 1983 dissertation;  we round off the portrait it by considering Gerard Debreu's impact on Yannelis' subsequent {\it oeuvre.}  In this connection, one can begin by singling  out  Debreu's 1952 {\it social existence theorem}: the concept of an abstract economy has its origin in  this pioneering paper.  Following,   \cite{ss75}, \cite{kv84}  and \cite{to84}, it gets taken up in \cite{ya87, ya09} and \cite{kpy}:  as delineated above, these papers  have  found application and extension in applied mathematics.  Hoewever, the issues are a little more subtle as far as applications in economics are concerned. The difference between an {\it economy} and a {\it game} is by now a staple of undergraduate courses: what is of  interest and not as equally  appreciated is the difference between  between an {\it abstract economy} in the sense of Debreu-Shafer-Sonnenschein and a {\it game.}   In his  Ely lecture, \cite{ar94}   goes into the difference. 
 \bqu  The current formulation of methodological individualism is game theory. In a game, each agent chooses one among a set of strategies available to him or her ... In the usual formulations, the set of available strategies is fixed, independent of the choices of others, and all the interactions among players are embodied in the payoff  functions. The choice of actions is totally individualistic. \equ
\noindent The point is that in an abstract economy, the actions are {\it not} independent: the very notion of an {\it abstract economy} was instrumentally motivated by the problem of the existence of a Walrasian equilibrium with each of  the individual action sets depending on the actions of the Walrasian auctioneer.  If the motivation is different, one has to work to show, as does \cite{re16c} for example,  that there are circumstances when an abstract economy can induce a games whose equilibria are the same as that of the economy, and hence is a suitable vehicle for applied work.\fn{Khan is grateful to Kalyan Chatterjee  who first discussed these issues as a room-mate at a conference in Bangalore in the early nineties. \cite{re16c} renames an {\it abstract economy} to be an {\it abstract game.} For a comprehensive discussion of these issues, in so far as they relate to the history of economic thought, see \cite{gku20}.}    It is then another interesting open problem  to apply the results obtained in Section 3.3 and 3.4 to investigate  the existence of an approximate equilibrium with externalities.\fn{The precursors for such an investigation are, for example,   \citet{so77, be92, sc15, os73, os77}, and the reader can begin with them  for hurdles  and difficulties in this line of research.}


\nocite{ce61, ce63}

\ifendnote
\pagebreak
\theendnotes
\fi

\section{Bibliography}

\nt This is a bibliography in two parts: the first provides detailed references to the chapter numbers cited in the text in keeping with the convention spelt out in Footnote~\ref{fn:convention};  the second collects the references to a technical and substantive economic literature cited in the text; 

%

\small{ 
\setstretch{1}


\subsection*{A~~~Selected Publications of Nicholas C. Yannelis}

\vspace{-8pt}


\bigskip 

\reff (1983) \lq\lq Existence of maximal elements and equilibria in linear topological spaces" (with
N. D. Prabhakar), {\it Journal of Mathematical Economics} 12,  233-245.

\medskip 

\reff -------- ``Existence and fairness of value allocation without convex preferences," {\it  Journal of Economic Theory}  31, 283-292.

\medskip 

\reff (1984) \lq\lq Non-symmetric cardinal value allocations" (with A. Scafuri), {\it Econometrica}  52, 136-1368 (Reprinted in the “Collected Papers of R. J. Aumann”, MIT Press, 2000)

\reff (1985) ``Maximal elements over non-compact subsets of linear topological spaces," {\it Economics Letters} 17, 133-136.

\medskip 

\reff -------- ``On a market equilibrium theorem with an infinite number of commodities," {\it Journal of Mathematical Analysis and Applications} 108,   595-599.

\medskip 

\reff -------- ``Value and fairness,” in {\it Advances in Equilibrium Theory}  C. D. Aliprantis, O. Burkinshaw and N. Rothman (eds.).  Berlin: Springer-Verlag.  

\medskip 

\reff -------- ``On perfectly competitive economies: Loeb economies,” (with D. Emmons), in {\it Advances in Equilibrium Theory}  C. D. Aliprantis, O. Burkinshaw and N. Rothman (eds.).  Berlin: Springer-Verlag. 

\medskip 

\reff (1986) ``Equilibria in Banach lattices without ordered preferences" (with W. Zame), {\it Journal of Mathematical Economics} 15, 75-110.

\medskip 

\reff (1987) ``Caratheodory-type selections and random fixed point theorems"  (withT. Kim and K. Prikry), {\it Journal of Mathematical Analysis and Applications} 122, 393-407.

\medskip 

\reff -------- ``Equilibria in non-cooperative models of competition," {\it Journal of Economic
Theory} 41, 96-111.

\medskip 

\reff
(1988) ``On a Caratheodory-type selection theorem" (with T. Kim and K. Prikry), {\it Journal
of Mathematical Analysis and Applications} 135,  664-670.

\medskip 

\reff -------- ``Fatou's lemma in infinite dimensional spaces," {\it Proceedings of the American 
Mathematical Society} 102, 303-310.

\medskip 

\reff (1989) ``Weak sequential convergence in $L_p(\mu,X)$," {\it Journal of Mathematical Analysis and
Applications} 141, 7283.

\medskip 

\reff -------- ``Equilibria in abstract economies with a measure space of agents and with an
infinite dimensional strategy space" (with T. Kim and K. Prikry), {\it Journal of
Approximation Theory} 56,  256-266.

\medskip 

\reff (1990) ``On the upper and lower semicontinuity of the Aumann integral," {\it Journal of Mathematical Economics} 19, 373-389

\medskip 

\reff (1991) “Equilibrium points of non-cooperative random and Bayesian games,” (with A. Rustichini), in {\it Riesz Spaces, Positive Operators and Economics,}  
C.D. Aliprantis, Border, K. and W. A. J. Luxemburg (eds.), Berlin:  Springer-Verlag.

\medskip 

\reff -------- ``On the existence of correlated equilibria” (with A. Rustichini), {\it Equilibrium Theory in Infinite Dimensional Space,}  M. Ali Khan and Nicholas C. Yannelis (eds.), Berlin: Springer-Verlag.

\medskip 

\reff -------- ``The core of an economy without ordered preferences,” {\it Equilibrium Theory in Infinite Dimensional Space,}  M. Ali Khan and Nicholas C. Yannelis (eds.), Berlin: Springer-Verlag.

\medskip 

\reff -------- ``What is perfect competition?” (with A. Rustichini), {\it Equilibrium Theory in Infinite Dimensional Space,}  M. Ali Khan and Nicholas C. Yannelis (eds.), Berlin: Springer-Verlag.

\medskip 

\reff -------- ``Set-valued functions of two variables in economic theory,” {\it Equilibrium Theory in Infinite Dimensional Space,}  M. Ali Khan and Nicholas C. Yannelis (eds.), Berlin: Springer-Verlag.

\medskip 

\reff -------- ``Equilibria in random and Bayesian games with a continuum of players,” (with E. Balder), in {\it Equilibrium Theory in Infinite Dimensional Space,}  M. Ali Khan and Nicholas C. Yannelis (eds.), Berlin: Springer-Verlag.

\medskip 

\reff -------- ``Equilibria in markets with a continuum of agents and commodities,” (with M. Ali Khan), in {\it Equilibrium Theory in Infinite Dimensional Space,}  M. Ali
Khan and Nicholas C. Yannelis (eds.), Berlin: Springer-Verlag.

\medskip 

\reff -------- ``The core of an economy with differential information," {\it Economic Theory} 1,  183-198.




\medskip 

\reff (1993) ``On the continuity of expected utility" (with E. Balder), {\it Economic Theory} 4,   625-643.

\medskip 

\reff -------- ``Commodity pair desirability and the core equivalence theorem,” (with A. Rustichini), in {\it General Equilibrium, Growth and Trade, II, The Legacy of Lionel W. McKenzie,}  M. Boldrin, R. Becker, R. Jones and W. Thomson (eds.), Academic Press. 


\medskip 

\reff
(1994) ``An elementary proof of the Knaster-Kuratowski-Mazurkiewitz-Shapley theorem"
(with S. Krasa), {\it Economic Theory} 4, 467-471.

\medskip 









\reff (1997)
``Existence of equilibrium in Bayesian games with infinitely many players" (with T. Kim), {\it Journal of Economic Theory} 77,  330-353. 

\medskip 







\reff (1998) “On the Existence of a Bayesian Nash Equilibrium,” in {\it Functional Analysis and Economic Theory,}  Y. Abramovich, E. Avgerinos, and N.C. Yannelis (eds.), Berlin: Springer-Verlag, 

\medskip 

\reff (2000) ``The Riesz-Kantorovich formula and general equilibrium theory"  (with C.D.
Aliprantis and R. Tourky), {\it Journal of Mathematical Economics} 34,  
55-76.

\medskip 

\reff -------- ``Cone conditions in general equilibrium theory" (with C. D. Aliprantis and R.
Tourky), {\it Journal of Economic Theory} 92,   96-121.

\medskip 





\reff (2001) ``A theory of value with non-linear prices: equilibrium analysis beyond vector
lattices" (with C.D. Aliprantis and R. Tourky), {\it Journal of Economic
Theory} 100,  22-72.



\medskip 

\reff -------- ``Markets with many more agents than commodities: Aumann's hidden
assumption" (with R. Tourky), {\it Journal of Economic Theory} 101, 
189-221.

\medskip 

\reff (2002) ``A Bayesian equilibrium existence theorem," {\it Advances in Mathematical
Economics} 4,  61-72.











\medskip 

\reff (2006) ``Continuity properties of the private core" (with E. Balder), {\it Economic Theory} 29,
 453 – 454. 



\medskip 

\reff (2007) ``Pure strategy Nash equilibrium in games with private and public information"
(with H. Fu , Y. Sun and Z. Zhang), {\it Journal of Mathematical Economics} 43,  523-531. 

\medskip 





\reff (2008) ``Saturation and the integration of Banach valued correspondences" (with Y. Sun),
{\it Journal of Mathematical Economics} 44,  861-865. 

\medskip 

\reff -------- ``Equilibrium theory with unbounded consumption sets and non-ordered
preferences, part I" (with D. Won), {\it Journal of Mathematical Economics} 44,
  1266-1283.





\medskip 

\reff -------- ``Equilibrium theory with asymmetric information and with infinitely many
commodities"  (with K. Podczeck), {\it Journal of Economic Theory} 141,
 152-183. 

\medskip 

\reff (2009) ``Debreu's social equilibrium theorem with a continuum of agents and
asymmetric information," {\it Economic Theory} 38,  419-432.

\medskip 

\reff -------- ``Bayesian Walrasian expectations equilibrium: beyond the rational expectations
equilibrium" (with E. Balder), {\it Economic Theory} 38,  385-397. 







\bigskip 

\reff (2011) ``Equilibrium theory with satiable and non-ordered preferences"  (with D. C.
Won), {\it Journal of Mathematical Economics} 47,  245–250. 











\medskip 

\reff (2014) ``Aggregate Preferred Correspondence and the Existence of a Maximin REE"   (with
A. Bhowmik and J. Cao), {\it Journal on Mathematical Analysis and Applications}  414,  29-45.

\medskip 

\reff -------- ``On the Existence of Mixed Strategy Nash Equilibria"  (with P. Prokopovych), {\it 
Journal of Mathematical Economics} 52,  87-97. 



\medskip 

\reff (2015) ``Discontinuous Games with Asymmetric Information: An extension of Reny’s
Existence Theorem" (with W. He), {\it Games and Economic Behavior} 91,   26-35. 

\medskip 



\reff (2016) ``Existence of Walrasian Equilibria with Discontinuous, Non-Ordered,
Interdependent and Price Dependent Preferences"  (with W. He), {\it Economic
Theory} 61,  497-513. 

\medskip 

\reff  --------`` Existence of Equilibria in Discontinuous Bayesian Games" (with W. He), {\it Journal of
Economic Theory} 162, 181-194. 

\medskip 

\reff  -------- ``A Remark on Discontinuous Games with Asymmetric Information and Ambiguity"  
(with W. He), {\it Economic Theory Bulletin}. 





\medskip 

\reff  (2017) ``Equilibria with Discontinuous Preferences: New Fixed Point Theorems" (with W. He), {\it Journal of Mathematical Analysis and Applications}  450,  1421-1433. 



\medskip 

\reff -------- ``On Strategic Complementarities in Discontinuous Games with Totally Ordered
Strategies" (with P. Prokopovych), {\it Journal of Mathematical Economics} 70,   147-173.









\medskip 

\reff (2019) ``On Monotone Approximate and Exact Equilibria of an Asymmetric First-Price Auctions with Affiliated Private Information" (with P. Prokopovych), {\it Journal of Economic Theory} (forthcoming). 

\bigskip



\setlength{\bibsep}{5.5pt}
\setstretch{1.12}

\bibliographystyle{../../References/econ} 
\bibliography{../../References/References.bib}

}
\end{document}